\documentclass[pdflatex,sn-nature]{sn-jnl}
\usepackage{float}
\usepackage{graphicx}%
\usepackage{multirow}%
\usepackage{amsmath,amssymb,amsfonts}%
\usepackage{algorithm}%
\usepackage{algorithmicx}%
\usepackage{algpseudocode}
\usepackage[all]{hypcap}
\usepackage{amsthm}%
\usepackage{booktabs}%
\usepackage{makecell}%
\usepackage[version=3]{mhchem}
\usepackage{setspace}%
\usepackage{placeins}

\makeatletter
\def\ps@myheadings{%
    \let\@oddfoot\@empty\let\@evenfoot\@empty
    \def\@evenhead{\thepage\hfil\slshape\leftmark}%
    \def\@oddhead{{\slshape\rightmark}\hfil\thepage}%
    \let\@mkboth\@gobbletwo
    \let\sectionmark\@gobble
    \let\subsectionmark\@gobble
    }
  \if@titlepage
  \renewcommand\maketitle{\begin{titlepage}%
  \let\footnotesize\small
  \let\footnoterule\relax
  \let \footnote \thanks
  \null\vfil
  \vskip 60\p@
  \begin{center}%
    {\LARGE \@title \par}%
    \vskip 3em%
    {\large
     \lineskip .75em%
      \begin{tabular}[t]{c}%
        \@author
      \end{tabular}\par}%
      \vskip 1.5em%
    {\large \@date \par}%
  \end{center}\par
  \@thanks
  \vfil\null
  \end{titlepage}%
  \setcounter{footnote}{0}%
}
\else
\renewcommand\maketitle{\par
  \begingroup
    \renewcommand\thefootnote{\@fnsymbol\c@footnote}%
    \def\@makefnmark{\rlap{\@textsuperscript{\normalfont\@thefnmark}}}%
    \long\def\@makefntext##1{\parindent 1em\noindent
            \hb@xt@1.8em{%
                \hss\@textsuperscript{\normalfont\@thefnmark}}##1}%
    \if@twocolumn
      \ifnum \col@number=\@ne
        \@maketitle
      \else
        \twocolumn[\@maketitle]%
      \fi
    \else
      \newpage
      \global\@topnum\z@
      \@maketitle
    \fi
  \endgroup
  \setcounter{footnote}{0}%
}
\makeatother

\begin{document}

\newcommand{\titlename}{Generative structural elucidation from mass spectra as an iterative optimization problem}

\title[]{\titlename}

\author[1]{\fnm{Mrunali} \sur{Manjrekar}}

\author[2,3]{\fnm{Runzhong} \sur{Wang}}
\author[1]{\fnm{Samuel} \sur{Goldman}}
\author[2]{\fnm{Jenna C.} \sur{Fromer}}

\author*[2,3]{\fnm{Connor W.} \sur{Coley}}\email{ccoley@mit.edu}

\affil[1]{\orgdiv{Computational and Systems Biology}, \orgname{Massachusetts Institute of Technology}, \orgaddress{\city{Cambridge}, \state{MA}, \country{USA}}}

\affil[2]{\orgdiv{Department of Chemical Engineering}, \orgname{Massachusetts Institute of Technology}, \orgaddress{\city{Cambridge}, \state{MA}, \country{USA}}}

\affil[3]{\orgdiv{Department of Electrical Engineering and Computer Science}, \orgname{Massachusetts Institute of Technology}, \orgaddress{\city{Cambridge}, \state{MA}, \country{USA}}}

\abstract{
  Liquid chromatography tandem mass spectrometry (LC-MS/MS) is a critical analytical technique for molecular identification across metabolomics, environmental chemistry, and chemical forensics. A variety of computational methods have emerged for structural annotation of spectral features of interest, but many of these features cannot be confidently annotated with reference structures or spectra. Here, we introduce FOAM (Formula-constrained Optimization for Annotating Metabolites), a computational workflow that poses structure elucidation from LC-MS/MS as an iterative optimization problem. FOAM couples a formula-constrained graph genetic algorithm with spectral simulation to explore candidate annotations given an experimental spectrum.
  We demonstrate FOAM's performance on the NIST'20 and MassSpecGym datasets as both a standalone elucidation pipeline and as a complement to existing inverse models. This work establishes iterative optimization as an effective and extensible paradigm for structural elucidation. 
}


\maketitle

\section*{Introduction}
\singlespacing

Liquid chromatography tandem mass spectrometry (LC-MS/MS) forms a critical component of the molecular discovery workflow present across many domains, including metabolomics \cite{Dang2009-tj, Watrous2012-bb, Quinn2020-qo, Li2022-ua, Klunemann2021-zb, Gentry2023-am}, environmental monitoring \cite{Strynar2015-uq, Hollender2017-cl, Tian2021-mo}, and chemical forensics \cite{Black1998-fr, Fraga2010-me, Roen2013-ue, Benito2015-ie}. The shared task across these domains is to elucidate the structure of a molecule from its experimental MS2 spectrum, commonly referred to as a ``fragmentation pattern'', in addition to a high resolution precursor mass that provides information about its molecular formula. Matching these experimental spectra to reference spectral libraries can provide annotations for the limited number of structures with reference spectra \cite{Schymanski2014-xy, Stein2012-ml, Wang2016-eh} but leaves the majority of spectra in complex samples unannotated \cite{Bittremieux2022-qs, El-Abiead2025-wx}. 

\begin{figure}
  \centering
  
  \includegraphics[width=\linewidth]{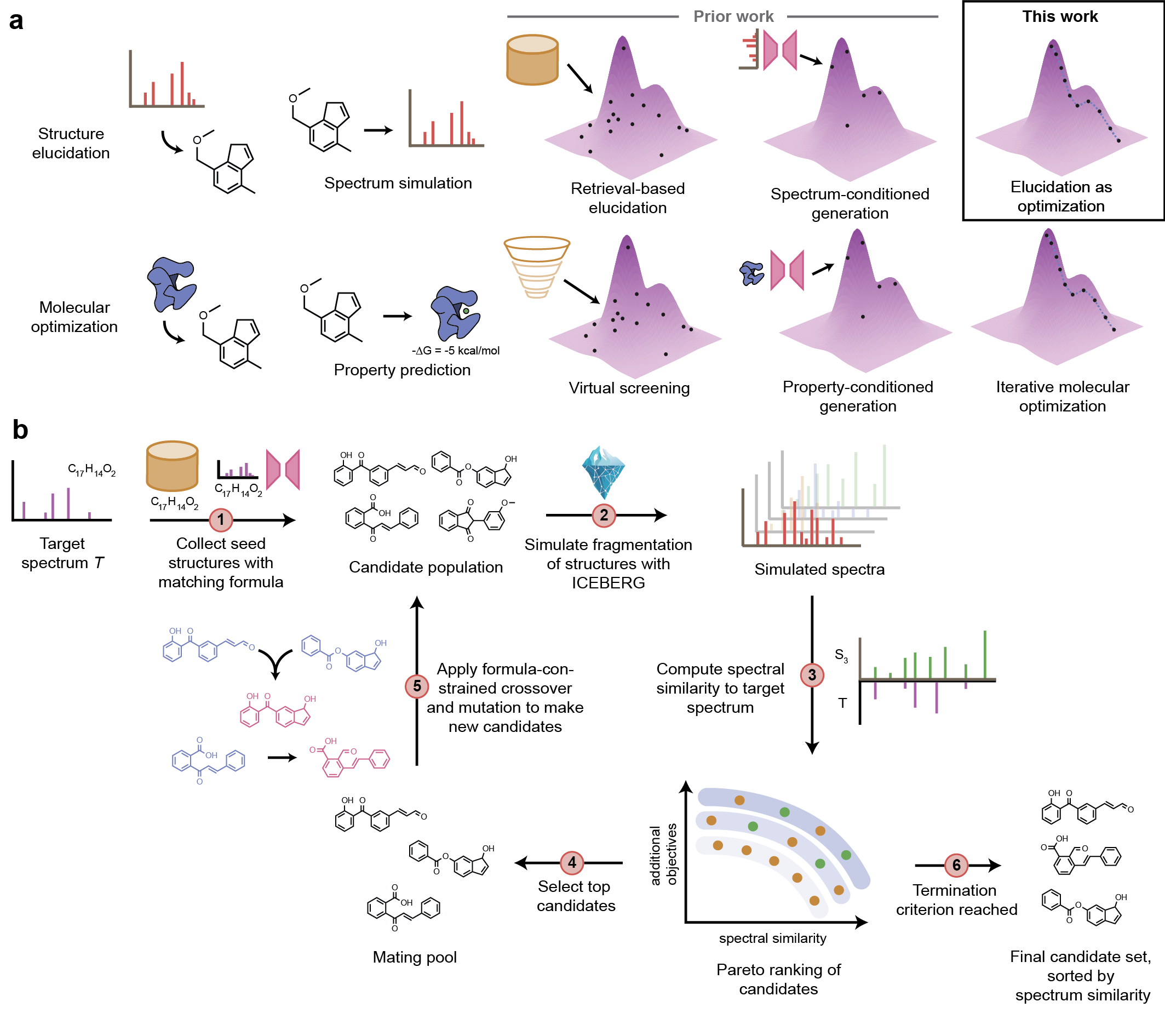}
  \caption{(a) We view structural elucidation as a special case of the molecular optimization problem common to computer-aided molecular design strategies; 
  our goal is to discover the molecule that best explains the experimental spectrum. A spectrum simulation model \cite{Wang2025-tc} and corresponding spectral similarity objective assume the role of a property prediction oracle. Spectral similarity is a shared pursuit of retrieval-based methods, which mirror virtual screening approaches in molecular design for identifying optimal candidates within a fixed list of candidates. Conditional generative models provide one option to propose new molecules \textit{de novo} for both drug design and structure elucidation \cite{Stravs2022-eo, DBLP:conf/icml/BohdeMWJC25}. Similarly, framing elucidation as an iterative optimization problem enables guidance by the spectral similarity objective in the same manner as iterative molecular optimization. 
  (b) Overview of the FOAM method. Given a target spectrum $T$, seed structures with matching formulae are first collected (for example, retrieved from a existing database or proposed by another elucidation tool) (Step 1). These structures are fragmented with a spectral simulator (Step 2), ICEBERG \cite{Goldman2024-nw, Wang2025-tc}, and compared to the target spectrum to compute their spectral similarities (Step 3). These similarities are considered alongside structural complexity (SAScore)~\cite{Ertl2009-rq} to calculate the Pareto ranking of candidates via non-dominated sorting. FOAM selects the top-scoring candidates to form the mating pool (Step 4) of parents for the next generation of offspring; FOAM then applies formula-constrained crossover and mutation operations to this pool to generate new candidates that maintain the desired formula (Step 5). The process repeats until the termination criterion is reached (Step 6), and the top-k candidates are extracted. }
  \label{fig:overview}
\end{figure}

Structural elucidation closely mirrors the more general challenge of molecular design (Figure \ref{fig:overview}a), as both endeavors pursue structures that optimize one or more objectives. In structural elucidation, the primary objective to pursue is the similarity between the experimental spectrum of interest and the spectrum of the hypothesized structure. Just as machine-learning and physics-based methods are applied to virtual screening, spectral simulation tools such as ICEBERG~\cite{Goldman2024-nw, Wang2025-tc}, CFM-ID~\cite{Wang2021-dw}, FraGNNet~\cite{Young2024-zp}, and others \cite{Murphy2023-te, Ruttkies2016-bl, Young2021-zi, DBLP:conf/icml/WangWMC25} can be used to rank untargeted compound libraries, such as PubChem~\cite{Kim2025-tg} by these structures' predicted spectral similarities to the observed spectrum. 

Enumerative strategies that apply chemical transformations \cite{Wishart2013-zd, Jeffryes2015-zd, Djoumbou-Feunang2019-ll, Mahjour2024-jp, Wang2025-tc} or produce constitutional isomers from molecular formula alone \cite{Lindsay1993-sv, Meringer2013-vq} can expand the considered chemical space beyond that captured by existing virtual libraries. Generative methods can, in principle, explore an even more diverse chemical space by learning to sample from the distribution of molecules in a reference library; two recent examples have demonstrated the untargeted generation of hypothetical metabolites given known ones \cite{Qiang2026-pa} and the untargeted generation of candidates with a particular molecular formula given enumerated constitutional isomers~\cite{Martin2025-cd}. 
Analogous to generative methods that design candidates conditioned on drug-relevant properties \cite{Tang2024-ne, Ozcelik2025-zq}, conditional generative models can learn a spectrum-to-structure mapping \cite{Stravs2022-eo, Litsa2021-po, Butler2023-pt, Wang2025-qh, DBLP:conf/icml/BohdeMWJC25}. These models are often probabilistic in nature, estimating the distribution $\Pr(\text{mol} \vert \text{spec})$ to sample a set of candidates when given a spectrum. 
Related strategies developed earlier have learned a spectrum-to-fingerprint mapping \cite{Duhrkop2019-qw, Goldman2023-tu} with the intention of querying against molecules in virtual libraries, rather than generating structures directly. Some of these recent conditional generative models leverage this learned fingerprint to condition their structure proposals. 
The dominant paradigm of computer-aided molecular design, arguably, is through iterative optimization. Surrogate models developed to screen virtual libraries can provide data-driven feedback to inform the design of new structures in a closed-loop manner. While the choice of the design protocol may vary---how molecules are proposed and how property scores inform the next round of design \cite{Fromer2023-ss, Du2024-jb}---the underlying goal is the progressive refinement of candidates through successive iterations. In structural elucidation, these iterative strategies could utilize spectral simulation surrogates to pursue the property of spectral similarity. Recent elucidation methods in the domain of NMR spectroscopy have pursued this optimization approach, adopting oracles that estimate spectral consistency (via simulation or latent representation) and graph-based generation mechanisms to produce candidates \cite{Mirza2024-jo, Jin2025-cf}. 

In this work, we describe FOAM (Formula-Constrained Optimization for Annotation of Metabolites), a method for \textit{de novo} structural elucidation from tandem mass spectra as an iterative optimization problem (Figure \ref{fig:overview}b). FOAM frames elucidation as a search constrained by a known chemical formula (as hypothesized or predicted from formula annotation tools). Its objective is to maximize predicted alignment to the experimental spectrum. FOAM employs ICEBERG \cite{Goldman2024-nw, Wang2025-tc}, a geometric deep learning model for spectral prediction, to fragment candidates \textit{in silico} and score their spectral similarities to the experimental spectrum. A formula-constrained and multi-objective graph genetic algorithm~\cite{Jensen2019-uz} iteratively proposes new structures for spectral similarity scoring that retain the specified molecular formula. These steps proceed iteratively until a specified termination criterion is met, and the final set of candidates are ranked by predicted spectral similarity. 

By framing structure elucidation from mass spectra as an optimization problem, FOAM provides several advantages. Foremost, it retains the spectral similarity objective inherent to retrieval-based methods and benefits from recent advances in accurate spectral prediction while not being limited to pre-enumerated candidate sets. In contrast to other spectrum-to-structure \textit{de novo} conditional generators, FOAM accepts any number of starting candidate structures to help orient its search. Finally, FOAM's multiobjective setup enables the optimization of auxiliary properties beyond spectral similarity alone. We demonstrate that FOAM achieves a high degree of success in recovering the true structure and ranking it highly (i.e., in its top 10 proposals) on a holdout set of the NIST'20 reference library, and demonstrates competitive benefits when utilized in tandem with other elucidation methods on the MassSpecGym benchmark.

\section*{Results}
\subsection*{The FOAM framework}
As an iterative optimization tool for elucidation, FOAM integrates (a) a structure-to-spectrum simulator, ICEBERG \cite{Goldman2024-nw, Wang2025-tc}, to predict spectral similarity to the reference spectrum with (b) a formula-constrained graph genetic algorithm to generate and modify hypothetical structure annotations (Figure \ref{fig:overview}b). 
We have recently demonstrated the utility of ICEBERG in quantitative benchmarks and prospective experimental case studies \cite{Wang2025-tc}, lending confidence to its adoption as a surrogate model in our optimization objective. In Step 1, given the target spectrum $T$ and corresponding MS1 information, formula annotators such as SIRIUS~\cite{Duhrkop2019-qw}, MIST-CF~\cite{Goldman2024-ay}, or BUDDY~\cite{ Xing2023-rz} can produce a chemical formula hypothesis $F$. We first retrieve a set of seed structures that match $F$ from a virtual library such as PubChem \cite{Kim2025-tg}. Conditional molecular generators and other elucidation tools can also contribute seed structures to this population, as discussed later. In Step 2, ICEBERG simulates these structures' spectra with the same instrumental parameters (collision energy, adduct, instrument if provided). In Step 3, we evaluate the spectral entropy similarities ~\cite{Li2021-vc} between the predicted spectra and the reference spectra. In Step 4, considering these similarities alongside auxiliary objectives (e.g., SAScore~\cite{Ertl2009-rq} for structural complexity in our present implementation), FOAM then constructs a Pareto ranking of structures and uses non-dominated sorting~\cite{Deb2002-se} to form the mating pool. 
From this pool, in Step 5, a graph genetic evolutionary algorithm \cite{Jensen2019-uz} applies formula-constrained crossover and mutation operations to generate new molecular candidates that maintain the desired formula. The steps of simulation, scoring, selection, and propagation constitute one generation in FOAM; these generations repeat until FOAM meets some termination criterion (Step 6). This termination criterion may be the number of generations, the number of scored molecules, or some percentage of molecules that pass a spectral similarity cutoff (absolute or relative). FOAM concludes by providing a candidate set sorted by spectral similarity. Importantly, FOAM only needs the experimental spectrum and target molecule's formula to make structure proposals.

We evaluate FOAM's ability to search for structures of interest on two commonly adopted benchmarks: NIST'20 \cite{NIST2020-nt} and MassSpecGym \cite{Bushuiev2024-jj}. For NIST'20, we adopt the random (seed=1) train-validation-test split as used by prior work \cite{Goldman2024-nw, Wang2025-tc}. In both settings, we train ICEBERG on the specified training set and perform elucidation experiments only on the test spectra. All elucidation experiments use a candidate population of 200 and generate up to 600 novel offspring in each generation. To regularize and improve the quality of structures, we incorporate the SAScore \cite{Ertl2009-rq} as a secondary objective penalizing structural complexity. We collect the top-1 and top-10 candidate sets ranked only by spectral similarity for evaluation. For NIST'20, seed structures include all PubChem formula matches (with the true structure excluded to simulate discovery), and FOAM terminates searches after 7,500 calls. For MassSpecGym, we additionally include samples from DiffMS \cite{DBLP:conf/icml/BohdeMWJC25} and terminate searches after 5,000 calls. We focus our primary evaluations on the NIST'20 set, as it is both larger and more uniform in reporting collision energies compared to MassSpecGym.

\subsection*{FOAM successfully optimizes candidate structures for spectral consistency}
\begin{figure}
  \centering
  \includegraphics[width=0.44\linewidth]{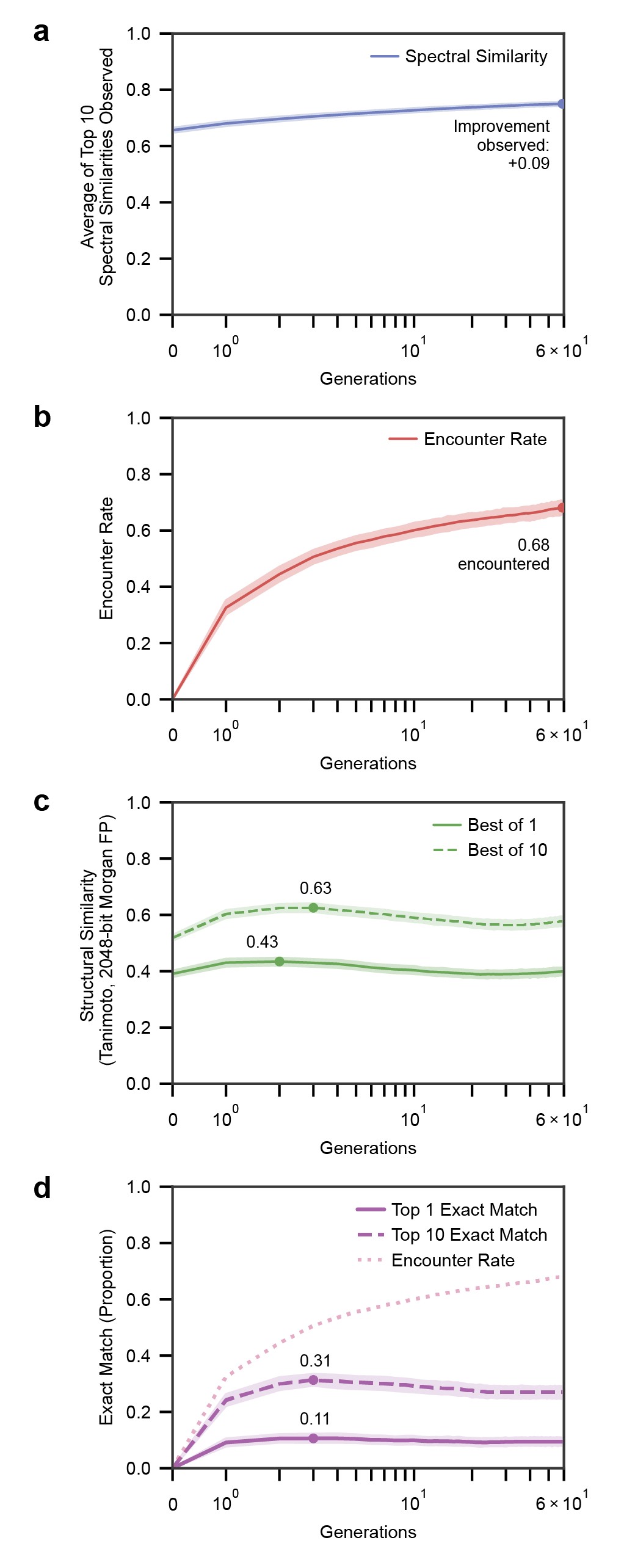}
  \caption{Performance of FOAM on a random test subset of NIST'20 as adopted from ICEBERG~\cite{Wang2025-tc}. (a) The primary objective function being maximized, the spectral similarity between the given experimental spectrum and the predicted spectrum of a proposed structure, exhibits a steady increase over the course of 60 generations. (b) FOAM proposes the true molecule 68\% of the time, irrespective of how the structure ranks among the full list of candidates. (c) Structural similarity, when taking only the top-scoring candidates as ranked by predicted spectral similarity, peaks at generation 2 (Best of 1) or 3 (Best of 10) and slowly decays afterward, reflecting the accumulation of competing decoy structures with higher predicted spectral similarity. (d) After only three generations, the true molecule is ranked first 11\% of the time and in the top 10 proposals 31\% of the time. The maximum values these can be are upper-bounded by the encounter rates in (b), as they are dependent on having seen the true molecule. Generations are plotted on a symmetric log scale. All metrics are plotted with 99.9\% confidence intervals, computed using bootstrapping (n=5,000) estimation of the mean.} 
  \label{fig:nist20}
\end{figure}

We first examine FOAM's performance as an optimizer, isolating its search efficacy from the accuracy of ICEBERG as a spectral prediction tool. 
The spectral similarity of the top candidates exhibits a consistent increase as the genetic algorithm proceeds through multiple generations (Figure \ref{fig:nist20}a); the average spectral similarity of the top 10 structures increases by 0.09 over the course of the search. This demonstrates that the formula-constrained mutation and crossover operations succeed in expanding the chemical space around seed molecules in a manner that enables incremental improvements each generation. 
Note that since this (averaged) metric represents predictive estimates of spectral consistencies, we do not expect these metrics to reach the hypothetical upper bound of 1.0. 

Our subsequent evaluations focus on FOAM's ability to encounter and highly rank the true chemical structure. With a sufficient number of generations allowed for exploration, FOAM encounters the true molecule two-thirds (68.1\%) of the time (Figure~\ref{fig:nist20}b), finding almost half (50.6\%) of these structures within the first three generations. Compared to the increase in spectral similarity we observe, the encounter rate exhibits a more dramatic increase, reflecting that the search remains productive even when spectral similarity improves slowly. 
The immediate increase observed after just one generation signals local refinement of seed structures; the subsequent steady increase demonstrates that further exploration of chemical space helps recover the true molecule. 
One reason for why the remaining 31.9\% searches fail to recover the true molecule is the imperfect nature of spectral prediction, which may lead the optimization toward unproductive or misleading structural changes. We examine how many of FOAM's searches find a molecule with equal or higher spectral similarity (which we call ``decoys'') (Supplementary Figure~\ref{fig:nist20_decoys_ms2match}a), and find that 95.2\% of searches eventually see at least one decoy. This demonstrates that FOAM is given an adequate computational budget to accomplish its primary optimization task in the vast majority of cases. Therefore, spectral prediction accuracy is the primary bottleneck to greater success rates rather than insufficient exploration of chemical space.

The encounter rate increases monotonically with the number of generations, by definition. However, as FOAM generates more high-scoring candidates, ranking the true structure highly among these candidates becomes increasingly challenging. To evaluate the top-1 and top-10 candidate sets as ranked by predicted spectral similarity, we consider their structural similarity to the true structure as measured by Tanimoto similarity (2048-bit, radius 2 Morgan fingerprint) (Figure~\ref{fig:nist20}c). Unlike the monotonic trends observed for spectral similarity and encounter rate, these performance metrics peak after 2-3 generations and drop afterward. These downward trends reflect the imperfections of the spectrum prediction oracle, which the genetic algorithm successfully exploits to generate decoys that outscore the true molecule.
The best structures in the top 10 candidate sets after two generations have on average structural similarities of 0.62 to the true structure. This is an improvement upon the similarity of the top-ranked candidates in the seed sets from 0.52 (Generation 0, akin to retrieving from PubChem and how ICEBERG has been applied previously). Further, adopting the criteria for ``meaningful'' and ``close'' matches introduced in Butler et al.~\cite{Butler2023-pt}, we show FOAM finds more meaningful matches and close matches in its proposals than in the starting seeds (Supplementary Figure~\ref{fig:nist20_decoys_ms2match}c-d).

We further illustrate FOAM's ability to suggest candidates that exactly match the true molecule in its top 1 and top 10 proposals (Figure~\ref{fig:nist20}d). These strict metrics require that FOAM both propose the true structure and rank it highly among the other many structural isomers. Note that these metrics are upper-bounded by the encounter rate. After 3 generations, FOAM ranks the true molecule in the top 10 proposals 31\% of the time, and as the top-ranked proposal 11\% of the time. These rankings on average tend to stay stable throughout the trajectory, worsening only slightly with extended generations. Given the challenging nature of the \emph{de novo} elucidation setting and the vastness of chemical space, this absolute performance is still notable. Altogether, these evaluations underscore FOAM's utility in exploring and prioritizing novel structures as possible annotations for LC-MS/MS spectra. 

\subsection*{FOAM's performance correlates with the relevance of seed structures and accuracy of spectral simulation} 
To probe which factors most strongly determine FOAM’s success, we analyze its performance along two key dimensions: the relevance of the starting seed structures and the accuracy with which ICEBERG predicts the fragmentation of the target molecule. We quantify these by analyzing FOAM's performance across both these contributing dimensions. Specifically, we first examine the structural similarity of FOAM's most-similar candidate among its top 10 proposals as a metric of success. We use a constant number of generations, three, which maximizes the Top 10 Exact Match recovery (Figure~\ref{fig:nist20}d).

\begin{figure}[t!]
  \centering
  \includegraphics[width=\linewidth]{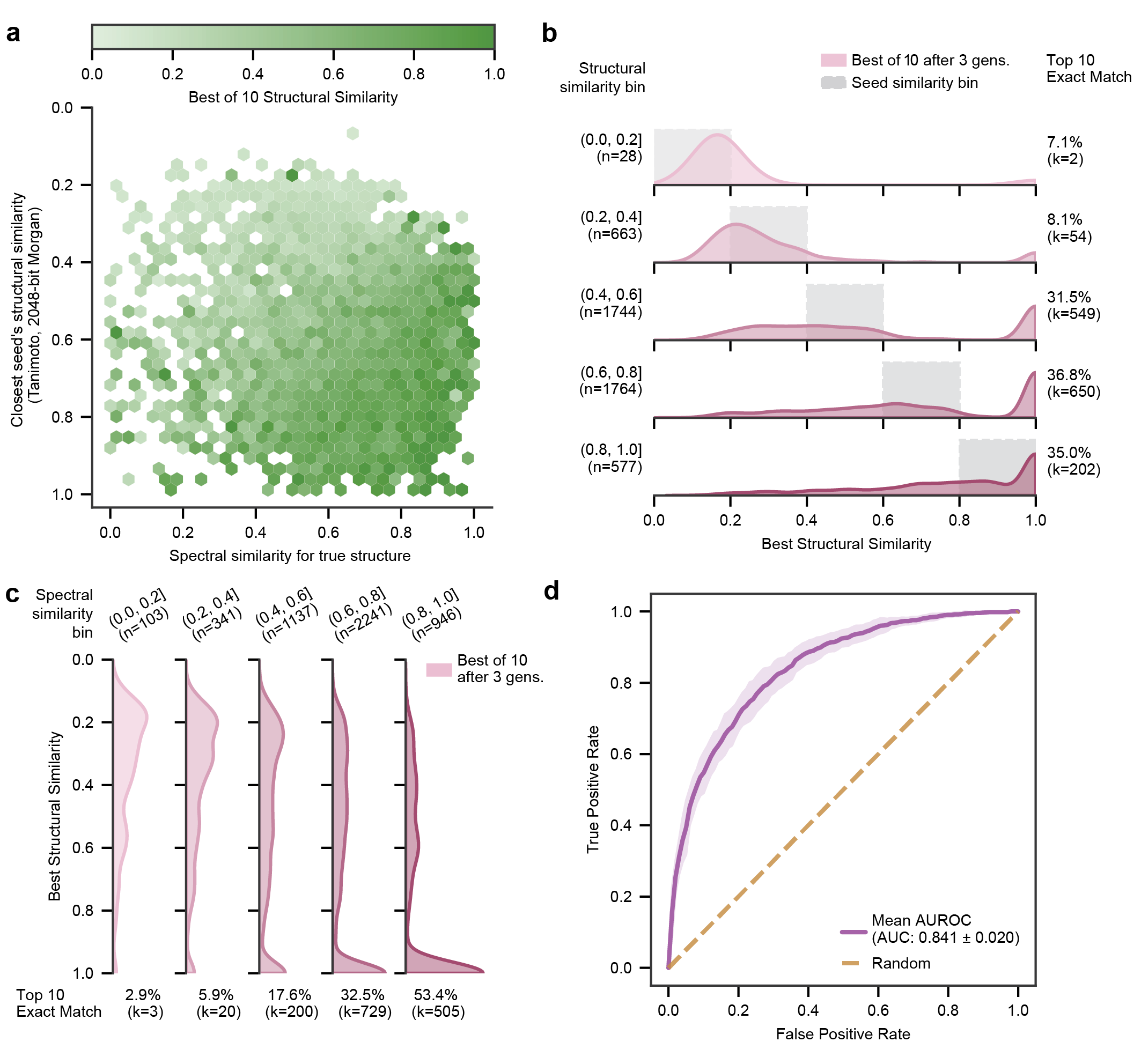}
  \caption{Contributions of seed similarity and spectral prediction accuracy to FOAM's performance, as defined by the composition of its top 10 candidate sets. (a) Best structural similarities observed in the top 10 candidate sets after three generations versus accuracy of the predicted spectrum of the true molecule (x-axis) and the similarity of the closest seed structure (2048-bit Morgan fingerprint, Tanimoto similarity) (y-axis). (b) Stratification of test spectra examples by starting seed similarity. Pink distributions show the structural similarity of the top 10 candidate sets after three generations. For visual reference, the seed similarity bins are also represented by corresponding gray range blocks for each density plot. The right-hand column quantifies the appearances of Top 10 Exact Matches in each bin. (c) Stratification of test spectra examples by spectral similarity of the true structure's predicted spectrum to the experimental spectrum. Note that since the density plots show the spread of \textit{structural} similarity, there are no corresponding gray range blocks to denote the bins. The bottom row quantifies the appearances of Top 10 Exact Matches in each bin. (d) Performance of a logistic regression model (5-fold cross-validation) predicting whether there is a Top 10 Exact Match in the final candidate set. The model achieves 0.841 AUROC using only information that would be available in prospective applications (maximum observed spectral similarity and its corresponding structure's SAScore, adduct type, count of collision energies acquired, generation count, and number of seed structures).} 
  \label{fig:limitations}
\end{figure}

As expected, we observe positive contributions toward the structural relevance of FOAM's top-ranked candidates from both the maximum structural similarity from the available seed structures (y-axis) and the similarity between ICEBERG's predicted spectrum and the observed spectrum of the true structure (x-axis) (Figure~\ref{fig:limitations}a). When the initial population contains structures close to the true structure, fewer modifications need to be made to arrive at the true one. When ICEBERG is able to accurately predict the true structure's spectrum, the objective function is more effective at guiding the search away from distracting incorrect structure hypotheses. Without structurally similar seeds and without an accurate spectral simulator, it is difficult for FOAM to rank structurally relevant compounds highly (upper-left quadrant). However, an accurate spectral prediction oracle mitigates cases where the starting population is highly dissimilar from the true structure (upper-right quadrant). 

We first isolate the benefit of having at least one seed that is structurally similar to the true structure, by binning examples based on their maximum structural similarity present among their seeds (Figure~\ref{fig:limitations}b). FOAM's probability of finding the true structure in the top 10 after 3 generations (``Top 10 Exact Match'') increases from 8.1\% to 31.5\% when the most-similar seed structure achieves a Tanimoto similarity of 0.4-0.6 instead of 0.2-0.4. Such a similarity threshold has been commonly found in other molecular optimization contexts to indicate structural closeness, such as distinguishing minimum (dis)similarity in clustering and downsampling workflows~\cite{Olivecrona2017-st, Mathai2021-nt}. 

Similarly, we stratify performance by how accurate ICEBERG is for the true structure, as measured by the predicted spectral similarity to its own experimental spectrum (Figure \ref{fig:limitations}c). Here, success steeply ascends with increasing spectral similarity; when ICEBERG predicts the spectrum for the true molecule with 0.8 or higher spectral similarity, the probability of finding the true molecule in the top 10 reaches 53.4\%. This trend underscores that the accuracy of the oracle is a critical determinant of FOAM's success. Indeed, when we allow FOAM 25 generations to explore (Figure \ref{fig:25gen_heatmap}), ultimately, only the accuracy of ICEBERG strongly correlates with the structural similarity of the candidate set. As improvements are made to ICEBERG or other spectral prediction models, those improvements should directly translate to increased success of \textit{de novo} elucidation with FOAM as well. 

The ability to estimate FOAM's confidence prospectively would inform decisions about how to allocate experimental resources when sourcing authentic standards. Yet seed similarity and simulation accuracy require \textit{a priori} knowledge of the true structure, and are thus unknown in prospective settings. To perform such prospective estimation of success, we train a logistic regression model to predict the Top 10 Exact Match at the end of the trajectory using features that do not require knowing the true structure. These features include information known about the spectrum itself, such as adduct type and number of collision energies collected, but also features from the trajectory, such as characteristics of the top candidates and number of generations elapsed. 

The logistic regression model achieves a mean AUROC of 0.841 in a five-fold cross-validation (Figure \ref{fig:limitations}d). The fit model exhibits positive coefficients for the spectral similarity and SAScore of the top-ranked candidate, as they both correlate positively with finding the true molecule in the top 10. Features describing the spectral metadata also contribute helpful discriminatory power. Adduct type can account for the variation in ICEBERG's retrieval accuracy (Figure \ref{fig:si_nist20_by_adducts}b), while higher counts of unique collision energies available can provide greater differentiation among candidates according to the spectral similarity objective. Together, these diagnostic signals form a practical assessment of confidence in 
estimating whether FOAM has succeeded in elucidating the true structure in experimental settings.

\subsection*{Optimization trajectories illustrate the evolution of top-ranking structure annotations}
\begin{figure}
  \centering
  \includegraphics[width=0.88\linewidth]{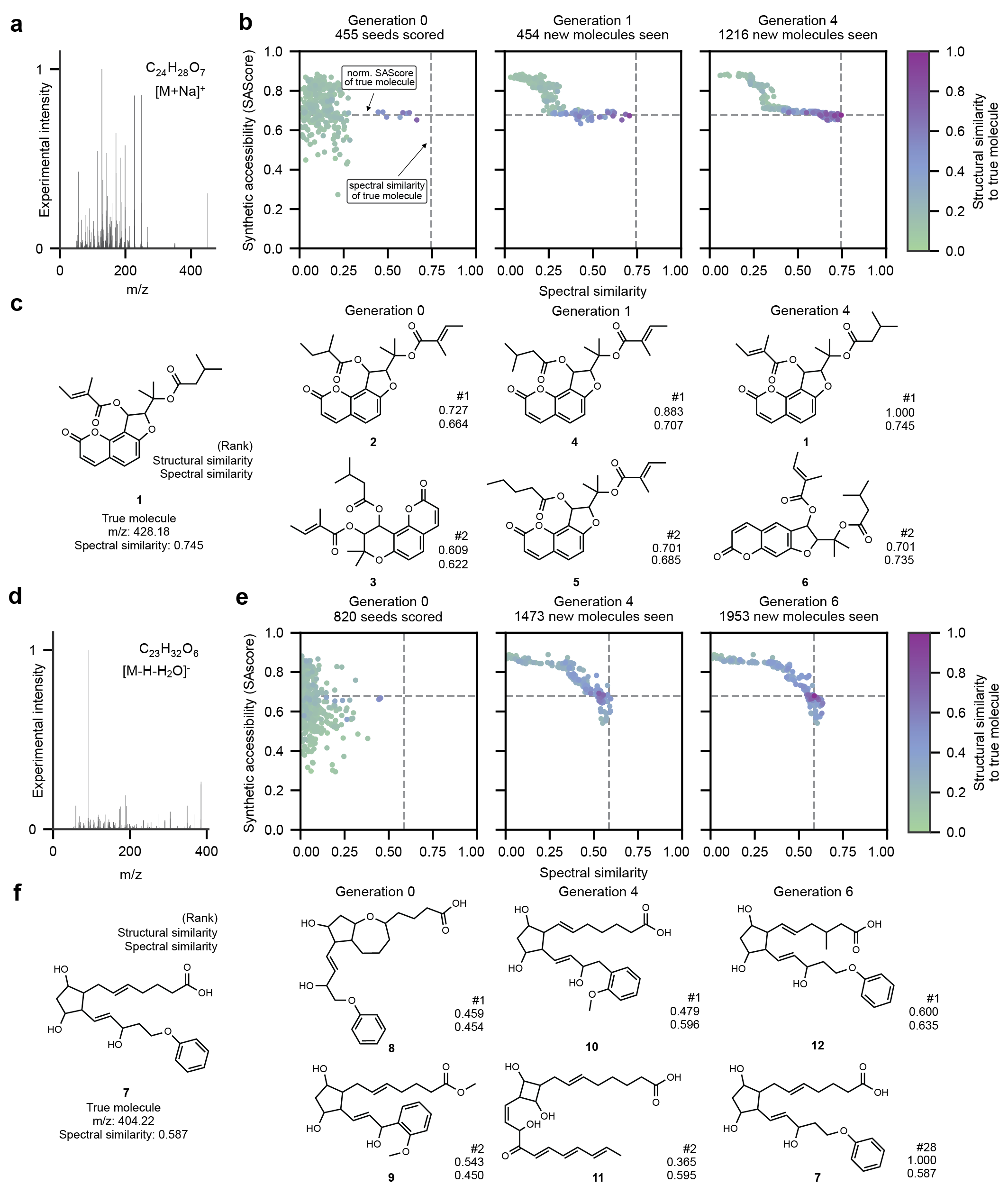}
  \caption{Two example FOAM trajectories and selected candidate structures for selected generations. (a) The molecular formula, adduct, and target spectrum merged over the 12 available collision energies (5\%, 9\%, 14\%, 20\%, 30\%, 40\%, 50\%, 65\%, 80\%, 95\%, 110\%, 130\%). (b) Distribution of objectives of the seed population (Generation 0), as well as the candidate populations after one and four generations. (c) The true structure \textbf{1} and the top two structures \textbf{2-6} from each generation ranked by spectral similarity. The true structure is found, and ranked first, in Generation 4. (d) The molecular formula, adduct, and target spectrum merged over the 11 available collision energies (7\%, 11\%, 15\%, 19\%, 25\%, 30\%, 36\%, 42\%, 50\%, 57\%, and 69\%).(e) Distribution of objectives as in (b) but for the search of \textbf{7}, with Generations 0, 4 and 6 visualized. (f) The true structure \textbf{7} and top-ranked candidates by spectral similarity. Generations 0 and 4 display their top two structures \textbf{8-11}; Generation 6 displays the top structure \textbf{12} and 28th structure. 
  The true molecule \textbf{7} is found in Generation 6 at rank 28, which is displayed in place of the 2nd ranked molecule for this generation. 
  Molecules are colored by their structural similarity (Tanimoto similarity, 2048-bit Morgan fingerprint) to the true molecule. Gray dashed lines mark the true molecule's ICEBERG-predicted spectral similarity to the experimental spectrum (vertical line) and the molecule's SAScore (horizontal line). 
  } 
  \label{fig:examples}
\end{figure}

We present two examples of elucidation trajectories 
in Figure~\ref{fig:examples}. 
We first examine a successful search (Top 1 Exact Match) for the metabolite \textbf{1}, whose spectrum, formula, and adduct we show in Figure~\ref{fig:examples}a. Figure~\ref{fig:examples}b displays selected snapshots of the distribution of the spectral similarities and SAScores of the populations at Generation 0 (seed structures from PubChem, excluding the true structure), and after 1 and 4 generations. For purposes of retrospective visualization, each point is colored by its structural similarity to the true molecule. The top two ranked structures by predicted spectral similarity from each generation are shown below (Figure~\ref{fig:examples}c). For this example, the vast majority of seed structures retrieved in Generation 0 have low spectral and structural similarity; no seeds have a spectral similarity that surpass that of the true structure (0.745) (Figure~\ref{fig:examples}b, Generation 0). However, there are a few structures with structural similarity exceeding 0.4, including \textbf{2} and \textbf{3} (Figure~\ref{fig:examples}c). Here, the top ranked seed structure (\textbf{2}) shares the true molecule's angular furanocoumarin scaffold. After one generation (Generation 1), selective pressure prunes the lower-scoring molecules in favor of the newly generated, higher-scoring molecules--the top two structures \textbf{4} and \textbf{5} now both share the same scaffold as the true structure. Though the top spectral similarity has only improved by 0.04, the Generation 1 snapshot reveals FOAM has generated numerous additional isomers with spectral similarities ranging from 0.5 to 0.7. By Generation 4, FOAM has found the true structure and ranks it as the top structure in the the population. Here, FOAM's rank two prediction (\textbf{6}) is a linear furanocoumarin instead of an angular furanocoumarin; all other connectivity is identical. 

We next examine the trajectory for an example where FOAM finds the true structure, a prostaglandin (\textbf{7}), but ranks it below the top 10 (Encountered, but no Top 10 Exact Match). The target spectrum for \textbf{7} (Figure \ref{fig:examples}d) exhibits relatively few high-intensity diagnostic peaks and is an [M-H+H\textsubscript{2}O]\textsuperscript{-} adduct, a poorly represented adduct in the NIST'20 training set (Table \ref{tab:nist20_adduct_training_exs}). Among the seed structures, only two structures (\textbf{8} and \textbf{9}) stand out with modest spectral similarity (0.45) to the experimental spectrum. Yet, these structures suffice to help the search progress forward toward more structurally useful candidates. In the Generation 4 snapshot, we see FOAM has produced many structures with spectral similarities that come close to that of the true molecule, with a number of candidates that have structural similarities of 0.6 or more. FOAM's top two ranked structures in this generation, \textbf{10} and \textbf{11}, are decoys with spectral similarities (0.596 and 0.595) that surpass that of the true molecule (0.587), though FOAM has not yet found the true molecule. FOAM only encounters the true molecule after 6 generations, but ranks it behind 27 other decoys encountered. The top-1 structure for this generation, \textbf{14}, shares reasonably high structural similarity (0.635), with only minor differences in connectivity to the true structure. 

\subsection*{FOAM is synergistic with other generative models for structure elucidation}
FOAM's use of a seed population of candidates to initialize its search makes it flexible for combination with candidate structures proposed by other elucidation models. We demonstrate such a use case with DiffMS \cite{DBLP:conf/icml/BohdeMWJC25}, a spectrum-conditioned generative model that uses discrete graph diffusion to propose candidates. We utilize the MassSpecGym benchmark \cite{Bushuiev2024-jj}, which features a challenging scaffold split but has been adopted for evaluation of many recent \textit{de novo} structural elucidation methods. For each spectrum in the MassSpecGym test set, we generate 100 samples from DiffMS, which may overlap in structural identity (Extended Data Figure \ref{fig:msg_evals}a) and provide them to FOAM as seeds alongside formula-matched structures from PubChem. Exact match exclusion is only applied to the PubChem seeds; any correct candidates generated by DiffMS remain in the seed populations. For fairness to other methods on the MassSpecGym benchmark, we retrain ICEBERG on the MassSpecGym training set and use this model in our optimization objective. 

Table \ref{tbl:main} shows the performance of DiffMS alone, DiffMS in combination with one generation of FOAM, and comparable methods. FOAM worsens DiffMS's Top 1 Exact Match accuracy, from 2.30\% to 1.50\%; however, these top-1 structures share higher structural similarity, improving the MCES (maximal common edge subgraph) edit distance from 13.96 to 12.21 and its Tanimoto similarity to the true structure from 0.28 to 0.35. Fewer Top 1 Exact Matches indicates that FOAM finds decoy structures with higher spectral similarity than that of the true structure. Yet, these top-ranking structures still share higher structural similarity to the true structure than DiffMS' best structures on average. FOAM also prioritizes many structurally relevant molecules in the top 10, making over twice as many successful true identifications (10.28\%) as DiffMS (4.25\%). These structures also show higher structural similarity to the true structure, sharing on average 0.53 Tanimoto similarity to the true structure. This is a considerable improvement over DiffMS alone, which only achieves 0.39 Tanimoto similarity, and reference methods. 

\begin{table}[!h]
  \caption{Structural elucidation performance on the MassSpecGym benchmark.}
  \label{tbl:main}
  \begin{tabular*}{\linewidth}{@{\extracolsep{\fill}}lcccccc}
  \toprule
  & \multicolumn{3}{c}{Top-1} & \multicolumn{3}{c}{Top-10} \\
  \cmidrule(lr){2-4}
  \cmidrule(lr){5-7}
  Model & Acc. $\uparrow$ & MCES $\downarrow$ & Tani. $\uparrow$ & Acc. $\uparrow$ & MCES $\downarrow$ & Tani. $\uparrow$ \\
    \midrule
     MIST + MSNovelist \cite{Stravs2022-eo}$^{*\ddag}$  & 0.00\% & 39.84 & 0.06 & 0.00\% & 18.83 & 0.15 \\
    MIST + Neuraldecipher \cite{Le2020-jz}$^{*\ddag}$  & 0.00\% & 22.93 & 0.14 & 0.00\% & 21.76 & 0.16 \\
    Random Generation \cite{Bushuiev2024-jj} & 0.00\% & 21.11 & 0.08 & 0.00\% & 18.26 & 0.11 \\
    MS-BART \cite{Han2025-bi} & 1.07\% &16.47 &0.23 &1.11\% &15.12 & 0.28 \\
    MADGEN \cite{Wang2025-qh} & 1.31\% & 27.47 & 0.20 & 1.54\% & 16.84 & 0.26 \\
    OMG (ESP) \cite{Martin2025-cd} & \textbf{2.42\%} & 55.43& 0.13 & 5.53\% & 54.21 & 0.18 \\  
    \hline
    DiffMS \cite{DBLP:conf/icml/BohdeMWJC25} $^{\ddag}$ & 2.30\% & 13.96 & 0.28 & 4.25\% & 11.68 & 0.39 \\
    FOAM (DiffMS + PubChem) & 1.50\% & \textbf{12.21} & \textbf{0.35} & \textbf{10.28\%} & \textbf{6.06} & \textbf{0.53} \\
\bottomrule
  \end{tabular*}
  \footnotetext{
  The Top 1 and Top 10 candidate sets from each method are evaluated for Exact Match accuracy (Acc.), maximum common edge subgraph edit distance (MCES), and Tanimoto similarity (Tani.). The Top 10 statistics reflect the best metric observed among the 10 suggestions (i.e., lowest MCES observed), not an average. Unless otherwise noted, all metrics are reproduced from corresponding works. $^{\ddag}$ indicates MCES was recalculated from what was originally reported. $^{*}$ indicates a method internally reimplemented, including reimplementations already shared in \cite{DBLP:conf/icml/BohdeMWJC25}. Myopic MCES is used to calculate the scores with a myopic threshold of 15, following the benchmark in \cite{Bushuiev2024-jj}. Tanimoto similarities are computed with 2048-bit radius-2 Morgan fingerprints. Recent methods where we could not confirm the absence of data leakage are excluded from this table.}
\end{table}

FOAM encounters the true structures in 33\% of the searches after just one generation, a considerable increase in coverage beyond the the 5.8\% of searches where DiffMS' samples (and thus FOAM seeds) already contained the true structure (Extended Data Figure \ref{fig:msg_evals}b). FOAM's top annotations also include higher proportions of ``Close Matches'' (Extended Data Figure~\ref{fig:msg_evals}b).  
 
\section*{Discussion}

In this work, we have formulated \textit{de novo} structural elucidation as an iterative optimization problem. Our method, FOAM, uses a formula-constrained graph genetic algorithm in combination with a forward spectrum simulation model. In contrast to other inverse models that operate as black-box translators, FOAM proposes and refines candidates based on their predicted spectral consistency with the target spectrum. In this manner, it approaches structural annotation as an iterative formula-constrained search through chemical space, starting from an initial candidate population provided by virtual libraries or complementary elucidation tools.

We demonstrate FOAM's ability to infer structural annotations on a holdout set of the NIST'20 reference library. FOAM encounters the true structure during its search in two-thirds of cases and succeeds in ranking the true molecule among its top 10 annotations 31\% of the time. Crucially, even when it does not rank the true structure first, FOAM's best structures nonetheless share high structural similarity to the true molecule, averaging 0.63 when considering the best structures among their top 10 annotations. 
By virtue of its core optimization objective, FOAM inherits the strengths and weaknesses of the specific spectrum simulator it uses. Here, our version of ICEBERG, trained on the NIST'20 reference library, supports a wide range of adducts, both positive and negative ionization modes, collision energy ranges, and structural classes; this coverage also extends to FOAM. We partially mitigate the risk of inaccurate spectral predictions by providing a post-hoc predictor of success from the test spectra trajectories, demonstrating a 0.841 AUROC in predicting the Top 10 Exact Match. 

Looking ahead, as FOAM's multiobjective framing can accommodate any number of objective functions to optimize simultaneously, we anticipate that further improvements to elucidation are possible with the incorporation of other orthogonal sources of information. Some examples of such contextual aids that FOAM could be modified to accept include metabolite-like classifiers \cite{Qiang2026-pa}, retention-time predictors~\cite{domingo2019metlin}, and measures of biosynthetic feasibility. 
Furthermore, while we develop methods to evaluate confidence post-hoc, if uncertainty was considered during the search itself, the current non-dominated sorting that determines the genetic algorithm's mating pool could be replaced with selection strategies from Bayesian optimization, perhaps improving the dynamics of FOAM's search. 

\pagebreak
\section*{Methods}
\subsection*{Details of FOAM's workflow and design choices}

\subsubsection*{Initialization with seed structures}
\label{seeding}
To begin its search, FOAM must be first provided a starting pool of candidate structures. These ``seed'' structures must match the chemical formula specified, which our experiments assume is known (for example, through successful inference from high-resolution MS1 in combination with the MS2 spectrum). At least one seed must be provided for the optimization to proceed. 
Seed structures are deduplicated and any stereochemical information is removed. In our experiments, our seed sets include (at least) all formula-matched molecules from PubChem as collected in January 2024, excluding the true molecule as determined by first layer InChI match. All seed structures are scored for their predicted spectral similarity to the true molecule. The scoring of these molecules does not count toward the maximum number of spectrum similarity evaluations when this is used as a termination criterion for the search.

\subsubsection*{Spectrum simulation and scoring}
To score candidate structures in terms of their agreement with the target spectrum, candidate molecular structures are fragmented with ICEBERG \cite{Wang2025-tc} under the same collision energies and adduct specified by the target spectrum provided. We then use the entropy similarity metric \cite{Li2021-vc} to quantify the spectral similarity between each of the candidate spectra and the target spectrum; benchmarking experiments with ICEBERG previously suggested that the entropy metric can be more informative than cosine similarity for ranking candidates \cite{Wang2025-tc}. When multiple collision energies are available, the entropy similarity is computed per individual collision energy, then averaged over all the energies to yield a single scalar measure of the similarity. Each spectrum has an assigned weight when averaging, such that spectra with more peaks, which are usually obtained at proper collision energies and are more informative, are considered more important. Any spectrum with 5 or more peaks has a weight of 4, 1-4 peaks a weight of 1, and 0 peaks a weight of 0. For simplicity throughout the text, we refer to the full set of target spectra corresponding to a single unknown structure as a singular \textit{spectrum}. The number of spectra per test instance varies from 1 to 29 in the NIST'20 dataset. In the MassSpecGym dataset, each test entry contains only a single collision energy, as the benchmark intentionally defines distinct collision energies for the same unknown molecule as distinct test examples. All molecules are stored with their scores in a buffer to avoid recomputation.

\subsubsection*{Structural complexity as a second objective}
The formula-constrained graph genetic algorithm can yield structures that are valid by chemical valence rules but highly unlikely to exist, e.g., due to unusual ring strain or rare substructures. 
To mitigate the risk that such unrealistic structures act as decoys and misguide the optimization trajectory, we incorporate a second objective of structural complexity in addition to spectral similarity. Specifically, we use the synthetic accessibility score (SAscore \cite{Ertl2009-rq}) as implemented in RDKit \cite{Landrum2024-km}, linearly normalized from 1-10 to 0-1:
$$S = \frac{10 - \text{SAScore}}{9}$$
so that a higher $S$ indicates a molecule that is \textit{easier} to synthesize. By incorporating this score as an second objective in a multi-objective optimization, FOAM can prioritize exploring candidates that are more structurally reasonable instead of structures that may have competitive spectral similarities but are too complex to be realistic annotations. 

\subsection*{Multi-objective optimization and Pareto ranking}
The spectral similarity and structural complexity objectives are incorporated into a commonly adopted multiobjective modification of genetic algorithms, NSGA-II ~\cite{Deb2002-se}. The NSGA-II framework relies on several key algorithmic choices:
\begin{itemize}
    \item \textbf{Non-dominated sorting} of candidates to produce a Pareto ranking of candidates. Molecules with the same Pareto rank form a front; within one front of candidates, no candidate dominates, or outcompetes on all objectives of, another candidate in that front. A Pareto rank of 1 describes a molecule that is not dominated by any other candidate in the population. 
    \item \textbf{Crowding distance} to sort candidates within a front; candidates are rewarded for having objective values distant from other values of its neighboring candidates in that front.
    \item \textbf{Binary tournament mating selection}, wherein candidates are chosen to mate based on whether it dominates another randomly selected candidate; this introduces a degree of stochasticity into which candidates are used to produce offspring for the next generation.
\end{itemize}
Empirically, we found that combining both objectives into a weighted, scalarized single objective problem tended to perform worse than the multiobjective approach. This is unsurprising given the general difficulty of balancing competing optimization criteria \emph{a priori}.
To identify the most performant molecules for propagation, we apply the non-dominated sorting described above to provide each molecule with a Pareto rank. Within a front, molecules are sorted by a objective-based crowding distance metric. In all experiment settings, we then select the top 200 structures to form the mating pool of the next population. 

\subsubsection*{Formula-constrained graph genetic algorithm}
Knowledge of the molecular formula allows us to constrain the generative space to produce only candidates that share the same chemical formula. We adapt the GraphGA algorithm \cite{Jensen2019-uz} as
implemented in PMO \cite{Gao2022-he} to constrain the crossover and mutation operators to only make modifications that preserve the formula. These operations are illustrated in \ref{sec:graph_ga_vis}.
The crossover operations swap fragments between two molecules with the same subformulae. These fragments must be different in connectivity to avoid the same molecules being recreated. To find these pairings, the order of bond is chosen (first, second, or third) at random. On both molecules, all potential bonds with this order are broken to enumerate a set of candidate fragments. Any pair of fragments that can be successfully be swapped to yield new molecules is chosen at random to form a new offspring. Similarly, mutations to the offspring generated can also be considered ``swaps'' between bonds on the molecule, i.e., moving any functional group connected by a single bond by swapping it with a hydrogen.

To begin the offspring creation process, we use binary tournament mating selection to construct pairs of parents from the mating pool. Specifically, each parent in a pair is set by picking two random candidates in the starting population and selecting the candidate with the higher Pareto rank (this selection occurs twice to set both parents in the pair). The crossover operation is applied first to create an offspring, and then a mutation operation is applied to that offspring to introduce a minor change with a default probability of 0.20. In the case of an initial seed population with fewer than 5 structures, crossover attempts are paused (since, in the case of only 1 structure, crossover cannot proceed, and with such few structures otherwise, empirically fails even more frequently) and multiple mutation attempts are performed instead to create a sufficient number of different offspring. In all experiment settings, 600 attempts at making new offspring are made in each generation. However, usually fewer than 600 new structures are made, either because the created offspring have already been seen before (in which case they are not added back to the population), because an operation results in an invalid molecule (as determined by RDKit's sanitization rules), or because no crossover is possible between a parental pair that would preserve the formula. 

\subsubsection*{Termination criteria}
The process of generation and scoring repeats until some termination criterion is reached. Furthermore, we apply a truncation mechanism that removes any structures with 0.35 spectral similarity or lower, once 20\% of the population is greater than 0.4 spectral similarity. This further encourages propagation of high-spectral-similarity scoring molecules, removing molecules that may be in the population only because of low structural complexity. 
Options for the termination criterion include reaching a maximum call threshold, a maximum generation threshold, or passing a minimum spectral similarity cutoff. For both experiment sets, we utilize a maximum call threshold (for NIST'20 experiments, it is 7,500 calls; for MassSpecGym experiments, it is 1,500 calls); seed structures do not count toward this threshold. Note that though each molecule must be evaluated separately by ICEBERG at each collision energy provided, we do not accumulate call counts across this dimension and only accumulate call counts across molecules. For all experiments, ranks reported are only based on spectral similarity.  
\subsubsection*{Summary of the FOAM algorithm}
An algorithmic framing of the overall method as applied for elucidation on the NIST'20 dataset is provided for clarity:
\begin{algorithm}
\caption{FOAM: Formula-constrained Optimization for Annotating Metabolites}
\label{alg:foam}
\begin{algorithmic}[1]
\Require target spectrum $\mathcal{T}$, molecular formula $F$, spectral simulator $\mathcal{O}$ (ICEBERG), seed library $\mathcal{V}$ (e.g PubChem), external elucidation tool $\mathcal{D}$ (e.g., DiffMS)
\Require max calls $N$, population size $k$, seed population size $s$, offspring size $\alpha$,  mutation size $\mu$, (optional, only for retrospective evaluation:) true structure $w$
\Ensure Ranked candidate set $\mathcal{C}$
\State $\mathcal{P}_0 \gets \textsc{RetrieveSeeds}(\mathcal{V}, F)$ \Comment{Formula-matched structures from library $\mathcal{V}$}
\If{D is defined}
\State $\mathcal{P}_0 \gets \mathcal{P}_0 \cup \mathcal{D}(T, F)$ \Comment{Additional suggestions from method $\mathcal{D}$}
\EndIf
\State $\mathcal{B} \gets \emptyset$ \Comment{Buffer of all scored molecules}
\State $t \gets 0,  g \gets 0$ \Comment{Call and generation counter}
\For{each molecule $m \in \mathcal{P}_0$} \Comment{Seed scoring}
        \State $\hat{s}_m \gets \mathcal{O}(m)$ \Comment{Predict spectrum}
        \State $\text{sim}_m, \text{sa}_m  \gets \textsc{EntropySimilarity}(\hat{s}_m, \mathcal{T}), \textsc{NormSAScore}(m)$ 
        \State $\mathcal{B}[m] \gets (\text{sim}_m,\text{sa}_m)$
    \EndFor
\If{the size of $\mathcal{P}_0$ is greater than $s$}
    \State $\mathcal{P}_0$ $\gets$ top $s$ matches from $P_0$ 
\EndIf
\State $\mathcal{P}_{1} \gets \mathcal{P}_{0}$ \Comment{Initialize next population}
\While{$t < N$}
    \For{$i = 1$ to $\alpha$} \Comment{\textit{Formula-constrained reproduction}}
        \State $\mathcal{M}_g \gets \mathcal{P}_{g}$ \Comment{Define mating pool}
         \State $(m_1, m_2) \gets \textsc{BinaryTournamentMating} (\mathcal{M}_{g})$ \Comment{NSGA-II selection}
         \State $m' \gets \textsc{Crossover}(m_1, m_2)$
         \State $m' \gets \textsc{Mutation}(m', \mu)$
         \If{$m' \notin \mathcal{B}$}
             \State $\mathcal{P}_{g+1} \gets \mathcal{P}_{g+1} \cup \{m'\}$
         \EndIf
    \EndFor
    \State $t \gets t + \lvert  \mathcal{P}_{g+1} \setminus \mathcal{B} \rvert$  \Comment{Increment call counter with num. of new molecules}
    \For{$m \in \mathcal{P}_{g+1} \setminus \mathcal{B}$} 
        \State $\hat{s}_m \gets \mathcal{O}(m)$
        \State $\text{sim}_m, \text{sa}_m  \gets \textsc{EntropySimilarity}(\hat{s}_m, \mathcal{T}), \textsc{NormSAScore}(m)$ 
        \State $\mathcal{B}[m] \gets (\text{sim}_m,\text{sa}_m)$
        \EndFor
    
    \State \Comment{\textit{Pareto ranking with NSGA-II and selection}}
    \State $\mathcal{R}_1, \mathcal{R}_2, \ldots  \mathcal{R}_{\lvert \mathcal{P}_{g+1} \rvert} \gets \textsc{NonDominatedSort}(\mathcal{P}_{g+1})$ \Comment{Sort by $(\text{sim}, \text{sa})$}
    \State $\mathcal{P}_{g+1} \gets \textsc{SelectK}(\mathcal{R}_1, \mathcal{R}_2, \ldots; k)$ \Comment{NSGA-II crowding distance to break ties}
    \State $g \gets g + 1$
\EndWhile
\State $\mathcal{C} \gets \text{Sort } \mathcal{P}_g \text{ by } sim_m \text{ descending}$
\State \Return $\mathcal{C}$
\end{algorithmic}
\end{algorithm}
\FloatBarrier

\subsection*{Details of quantitative evaluations}
\subsubsection*{NIST'20 benchmarking}
For the NIST'20 experiments, we use a version of the ICEBERG model trained on a random (2D InchiKey disjoint) split of the NIST'20 training set, following the same data split as ICEBERG~\cite{Wang2025-tc}. This dataset features 10 adducts ([M+H]\textsuperscript{+}, [M\textminus H]\textsuperscript{\textminus}, [M\textminus H\textsubscript{2}O+H]\textsuperscript{+},[M+Na]\textsuperscript{+}, [M\textminus 2H\textsubscript{2}O+H]\textsuperscript{+}, [M\textminus H\textminus CO\textsubscript{2}]\textsuperscript{\textminus}, [M\textminus H\textminus H\textsubscript{2}O]\textsuperscript{\textminus}, [M+NH\textsubscript{3} +H]\textsuperscript{+},[M+Cl]\textsuperscript{\textminus}, [M+K]\textsuperscript{+}), whose training counts can be found in Supplementary Table \ref{tab:nist20_adduct_training_exs}. We therefore only evaluate FOAM's performance on spectra that belong to the test fold of this split (4,925 test instances) to avoid concerns of data leakage. Searches are run for a maximum of 7,500 calls to ICEBERG. 

\subsubsection*{Confidence prediction model}
To estimate the probability of FOAM's elucidation success in prospective settings, we build a binary classifier to predict whether the true molecule will appear in the Top 10 candidate set (``Top 10 Exact Match''). We focus on predicting this outcome at the end of the trajectory (7,500 calls), where approximately 28\% of cases have a true molecule in the Top 10. We train a logistic regression model (from the scikit-learn library \cite{scikit-learn}) on the following features: 
\begin{enumerate}
    \item Adduct type of the spectrum (one-hot encoded; 10 possible adducts)
    \item Number of collision energies available for the test structure (range: 1 to 30)
    \item Normalized SAScore of the best identified molecule (ranked by spectral similarity)
    \item Spectral similarity of the best identified molecule (ranked by spectral similarity)
    \item Generation count
    \item Number of seed structures
\end{enumerate}
For model evaluation, we employ a 5-fold cross-validation strategy on the dataset. The data is partitioned into five subsets; in each iteration, a logistic regression model is trained on 80\% of the data and evaluated on the remaining 20\% hold-out fold. We standardize all features, Each logistic regression model is trained with binary cross-entropy loss, L1 regularization (C=1.00), and balanced class weighting. For each model, we average the true positive rates across all hold out sets, then compute the Area Under the Receiver Operating Characteristic curve (AUROC) to get an averaged AUROC, which we use as our primary evaluation metric. 

\subsubsection*{MassSpecGym benchmarking}
\label{sec:msg_benchmark_details}
We further benchmark on the MassSpecGym dataset~\cite{Bushuiev2024-jj}. This open-access dataset features 231,104 spectra; we use the provided data split, which emphasizes generalizability, as each test structure has at least an MCES (edit distance) of 10 from any training structure. We retrain a version of ICEBERG that respects this same data split, the details of which are found in \ref{sec:si_msg_benchmark_details}. For each of the test spectra in the MassSpecGym benchmark, we seed structures with both formula-matches from PubChem (following the same protocol as for NIST'20) and all formula-matched candidates generated by DiffMS \cite{DBLP:conf/icml/BohdeMWJC25}. Up to 100 proposals are sampled from DiffMS for each spectrum. Since DiffMS is heavy-atom-constrained, we filter out any proposals that do not exactly match the specified chemical formula. We exclude the true structure from the seeds retrieved from PubChem (as identified by first layer InChI match), but retain it if it appears among the DiffMS suggestions. Searches are run for a maximum of 1,500 ICEBERG calls.

\subsubsection*{Excluded baseline methods}
Comparisons and potential extensions of some recently released \textit{de novo} generation models, namely MIST+MolForge ~\cite{Neo2025-wj}, Gaia-01 \cite{Novogaia2025-ez}, and TTT \cite{Mismetti2025-ze} are excluded as we could not confirm the absence of data leakage in their reported implementation. 
Given that FOAM successfully improves over DiffMS-generated structures, as shown in Table~\ref{tbl:main}, we expect FOAM to serve as an orthogonal approach to improve any other \textit{de novo} generation methods.

\subsection*{Evaluation metrics}
To evaluate our Top 1 and Top 10 candidate sets, which are extracted from all candidates ranked by spectral similarity only, we use "Best of 1" to refer to our Top 1 rank structure's metric, and "Best of 10" to refer to the best of the top 10 ranked structures' metrics. 
\begin{itemize}
    \item Spectral similarity: spectral entropy similarity \cite{Li2021-vc}
        \begin{equation}
        \text{specsim}(\mathcal{T}, \hat{s}) = 1 - \frac{2\,H(0.5 (\mathcal{T} + \hat{s})) - H(\mathcal{T}) - H(\hat{s})}{\ln(4)}
        \end{equation}
    \item Tanimoto similarity: Let $FP$ be our 2048-bit Morgan fingerprint (radius 2) generator from RDKit (version 2023.9.6), then:
        \begin{equation}
        \text{Tani}(m_1, m_2) = \frac{|FP(m_1) \cap FP(m_2)|}{|FP(m_1)| + |FP(m_2)| - |FP(m_1) \cap FP(m_2)|}
        \end{equation}
        \begin{itemize}
            \item \text{Best of 1: } $T(w, m_{(1)})$
            \item \text{Best of 10: } $\max_{i \in \{1 \dots 10\}} T(w, m_{(i)})$
        \end{itemize}
        where $w$ is the ground truth structure.
    \item Exact Match: Given the first layer (connectivity) of 2 InchiKeys (first 14 characters), we measure Exact Match by measuring whether they equal each other:
    \begin{equation}
        \text{Exact Match}(m_1, m_2) = (\text{2D InchiKey}(m_1) == \text{2D InchiKey}(m_2))
        \end{equation}
    \item Maximum common edge subgraph edit distance \cite{Kretschmer2025-of}: 
        \begin{equation}
            \text{MCES}(G_{m_{1}}, G_{m_{2}}) = |E_{m_{1}}| + |E_{m_{2}}| - 2|E_c|
        \end{equation}
        \begin{equation}
            G_c = (V_c, E_c) \subseteq G_{m_{1}}, G_{m_{2}} \text{ s.t. } |E_c| \text{ is maximized}
        \end{equation}
        \begin{itemize}
        \item \text{Best of 1: } \text{MCES}($w, m_{(1)}$)
        \item \text{Best of 10: } $\min_{i \in \{1 \dots 10\}} \text{MCES}(w, m_{(i)})$
        \end{itemize}
    
\end{itemize}
\setcounter{figure}{0}
\renewcommand{\figurename}{Extended Data Fig.}
\renewcommand{\theHfigure}{Extended Data Fig.\thefigure}

\setcounter{table}{0}
\renewcommand{\tablename}{Extended Data Table}
\renewcommand{\theHtable}{Extended Data Table\thetable}
\backmatter

\bmhead{Acknowledgements}
We thank Joules Provenzano, Wenhao Gao, Jihye Roh, Babak Mahjour, and Magdalena Lederbauer for assistance and feedback during development. 

\section*{Declarations}

\textbf{Funding:} This work was supported by DSO National Laboratories in Singapore and the MIT Generative AI Impact Consortium (MGAIC). M.M. was supported by NIH grant T32GM087237 and by an NSF Graduate Research Fellowship (grant no. 2141064).

\noindent\textbf{Conflict of interest:} The authors declare no competing interests.

\noindent\textbf{Ethics approval and consent to participate:} Not applicable.

\noindent\textbf{Code availability:} FOAM is open-source; its code, documentation, and analytical scripts can be found at \url{https://github.com/coleygroup/foam}. Configuration files are provided, as are the seed structure caches, to reproduce all experiments performed in the paper and to facilitate straightforward modification for its usage in other structural elucidation campaigns. 

\noindent\textbf{Data availability:} Code and data are available at \url{https://github.com/coleygroup/foam}. The NIST 2020 dataset is under commercial license and cannot be freely redistributed. The MassSpecGym dataset may be found at \href{https://huggingface.co/datasets/roman-bushuiev/MassSpecGym/}{https://huggingface.co/datasets/roman-bushuiev/MassSpecGym/}. MassSpecGym model weights, as well as PubChem and DiffMS seed structure caches, may be found on Zenodo: \url{doi:10.5281/zenodo.18502041}.

\textbf{Author contributions:} C.W.C., M.M., and S.G. conceptualized the project. M.M., R.W., and J.F. implemented the algorithmic framework. M.M. and R.W. trained models. M.M. ran and analyzed the experiments. All authors contributed to the writing of the manuscript. 

\bibliography{sn-bibliography}

\begin{figure}[t!]
  \centering
  \includegraphics[width=\linewidth]{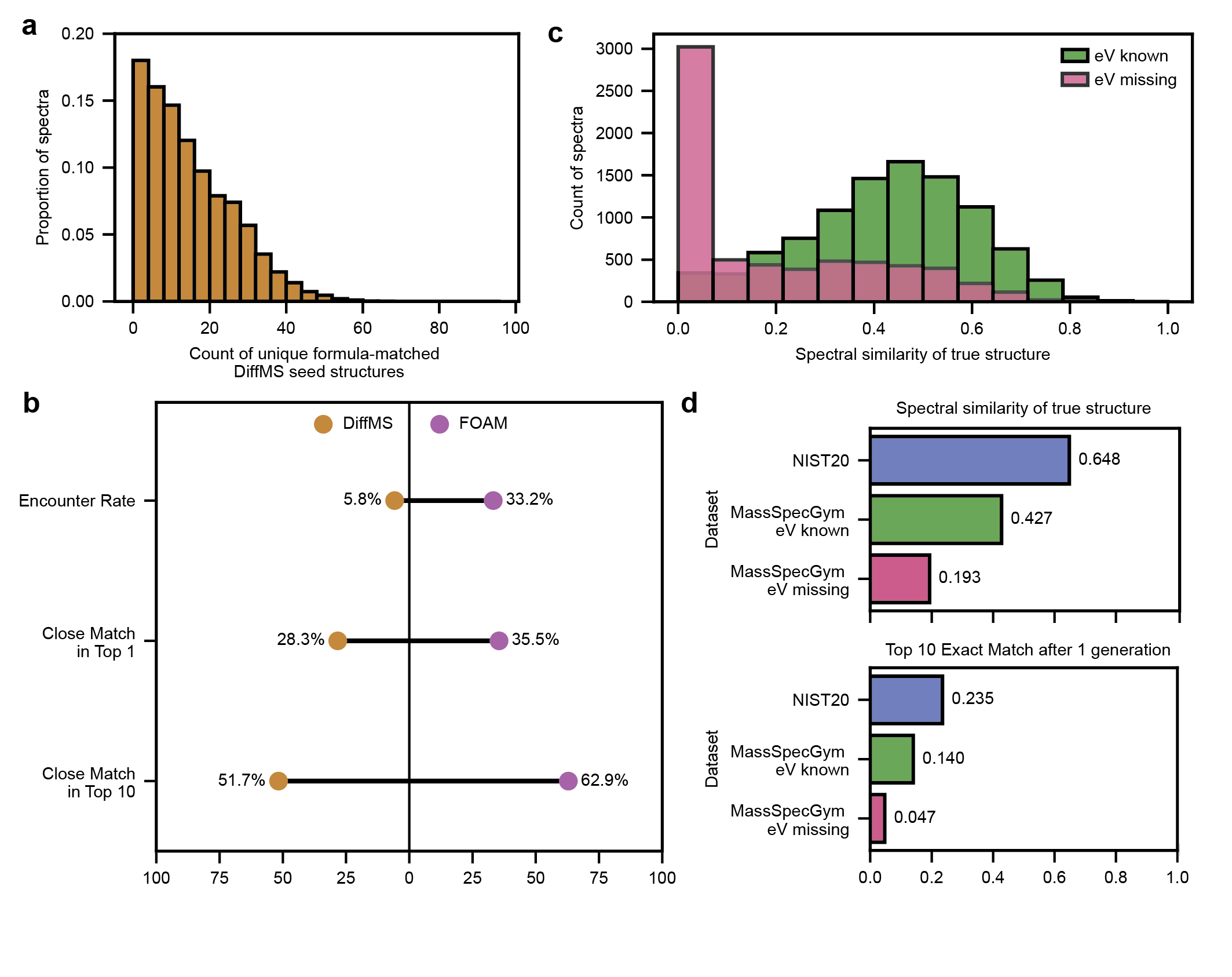}
  \caption{Additional evaluations of FOAM on the MassSpecGym dataset. (a) Histogram of unique samples collected from DiffMS; 100 (possibly overlapping) samples were collected per spectrum. (b) Distribution of simulated spectral similarity for the true structure for spectra in the MassSpecGym dataset where collision energy (eV) was either already annotated (known) or was missing (missing). (c) Comparison between statistics from DiffMS to FOAM for encounter rate, and count of close matches in the top 1 and top 10 candidate sets as defined by \cite{Butler2023-pt}. (d) Comparison of spectral similarity averages and Top 10 match rates for NIST'20 vs MassSpecGym, which are dramatically lower when the collision energy needed to be imputed. }
  \label{fig:msg_evals}
\end{figure}

\clearpage

\setcounter{figure}{0}

\renewcommand{\theHfigure}{Supplementary Fig.\thefigure}
\renewcommand{\figurename}{Supplementary Fig.}
\renewcommand{\thefigure}{S\arabic{figure}}
\setcounter{table}{0}

\renewcommand{\theHtable}{Supplementary Table\thetable}
\renewcommand{\tablename}{Supplementary Table}
\renewcommand{\thetable}{S\arabic{table}}
\renewcommand{\theHsection}{Supplementary Section\thesection}
\renewcommand{\thesection}{S\arabic{section}}
\setcounter{section}{0}

\clearpage
\setcounter{page}{1} 
\title{Supplementary Information for \titlename}
\abstract{\null}
\renewcommand{\abstractname}{}

\maketitle
\clearpage
\tableofcontents
\clearpage

\section{FOAM implementation details}
\subsection{Seeding cache}
Candidates are collected from a local cache of PubChem as downloaded in January 2024. Chemical formulas were computed for all structures after deduplication across tautomers and stereoisomers. Salts, fragmented structures, and molecules heavier than 1,500 Dalton are excluded from this cache to follow the constraints of ICEBERG.

\subsection{Crossover and mutation operations}
\label{sec:graph_ga_vis}
We provide visual examples of the crossover and mutation operations applied in Supplementary Figure \ref{fig:crossover_mutations}. In panel (a), we show non-ring-based crossover between two parents with both formulas of \ce{C16H12O3}. We show an example of a single-bond breakage found in both parents that yields two fragments per parent, each of which has a corresponding formula-matched fragment in the other parent. That correspondence is used to define the swap; here, the \ce{C9H7O} fragments swap places, the bonds are reconstructed to create two new molecules. One of these molecules is chosen at random (50\% probability) to be the offspring of this parent. Further, we also illustrate ring crossover, which now focuses on finding and applying the same swap to ring-based bonds. In panel (b), we show examples of mutation operations that we cover. Any mutation can be posed as a swap of connectivity or shift of electrons. The mutation operations we offer include swaps between a hydrogen and a single bond (first row), as well as swaps between double and triple bonds (not pictured); ring breaking or formation (second row); changes in bond orders to two places on the molecule (third row). The ring formation/breaking and bond order changes can also be crossed together, i.e. where a ring forms, an saturation takes place elsewhere on the molecule to accommodate the freed hydrogens. Additionally, our mutation operations also cover moving a heteroatom (e.g. O, N) around a ring system (fourth row); and swapping multiple single-bonds for one double bond (fifth row). There is also a ring fusion operation that helps unite two closely joined rings into a fused system (not pictured). 
We group these into 7 different mutation operations; for every offspring selected to mutate (default probability 20\%), Supplementary Table \ref{tab:si_mutation_type_probs} details the probabilities associated with each operation. If there are no rings in the molecule to be mutated, the probabilities of operations involving rings (operations 2, 3, 4, 6, 7) are set to 0 and the remaining probabilities are normalized to sum to 1. Similar normalization occurs also if there are no two neighboring rings, which is necessary for the ring fusion operator (operation 7) to be applied. 

\begin{figure}[ht!]
  \centering
  \includegraphics[width=\linewidth]{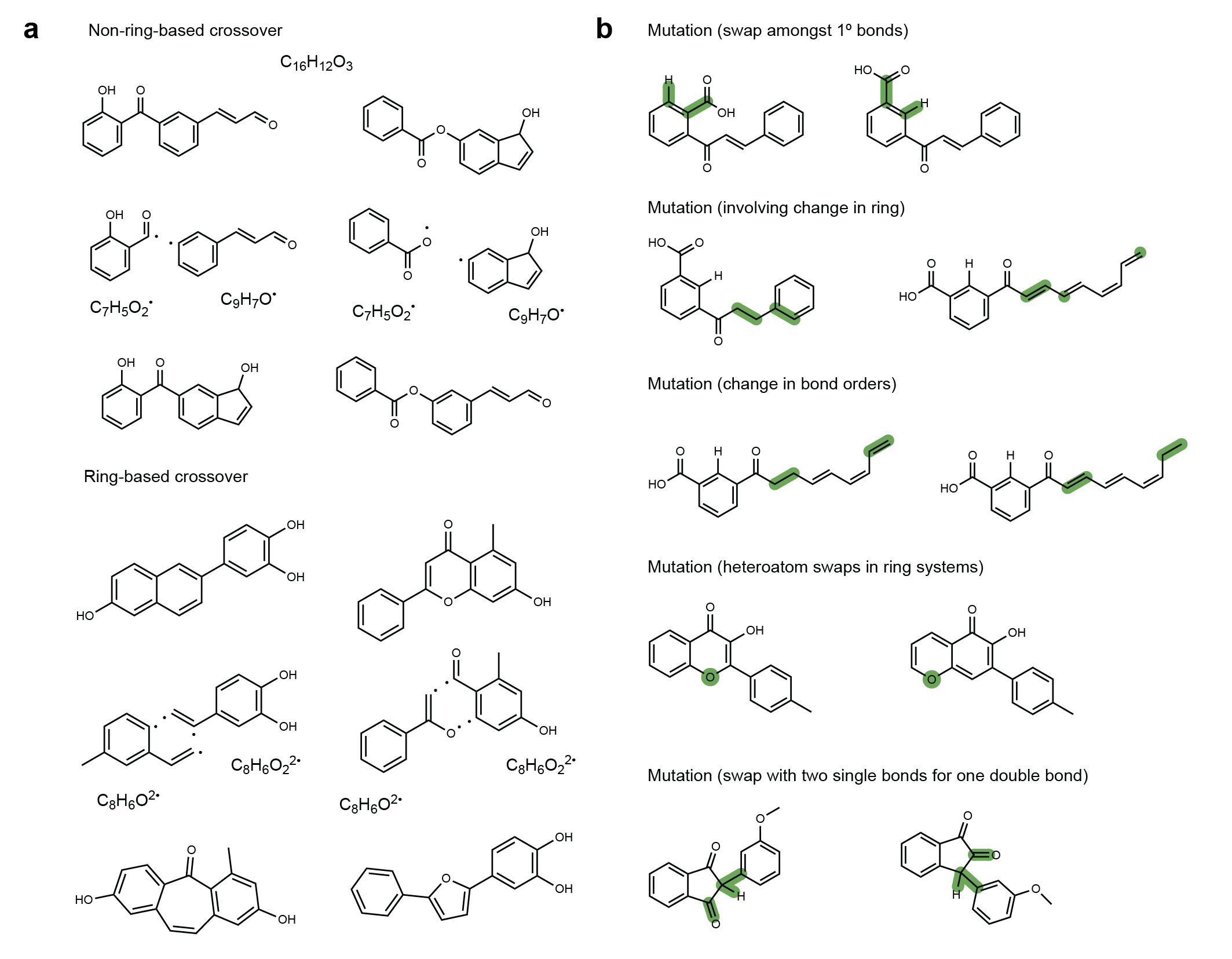}
  \caption{Visualization of examples of formula-constrained crossover and mutation. (a) Crossover examples, through cuts made to non-ring bonds (top) or ring-based bonds (bottom). Fragments made must match in chemical formula for crossover to proceed. (b) Overview of possible mutations that preserve formula. }
  \label{fig:crossover_mutations}
\end{figure}

\begin{table}[h!]
    \centering
    \begin{tabular}{c|c|c}
    \toprule
        Number & Mutation operator & $p$ \\
        \midrule
        1 & Bond order upgrade and downgrade & 0.14 \\
        2 & Bond order downgrade, ring formation & 0.07 \\
        3 & Bond order upgrade, ring break & 0.07 \\
        4 & ring formation, ring break & 0.16 \\
        5 & Cut and paste swaps & 0.28 \\
        6 & Heteroatom migration & 0.14 \\
        7 & Ring fusion & 0.14 \\
    \bottomrule
    \end{tabular}
    \caption{Probabilities associated with each mutation operator type. Operations are visualized in \label{sec:graph_ga_vis}.}
    \label{tab:si_mutation_type_probs}
\end{table}

\clearpage

\subsection{Extended benchmarking details}
\subsubsection{NIST'20 evaluation}
For the NIST'20 experiments, we use a model trained on a random (2D InchiKey disjoint) split of the NIST'20 training set, consisting of 429,751 training spectra covering 20,676 molecules and 53,074 testing spectra covering 2,556 molecules with random seed=1. This follows the same data split as ICEBERG~\cite{Wang2025-tc} to make sure there is no data leakage; all training details for this model may be found in the corresponding citation. Spectra are separated by molecule and adduct, but grouped by collision energy, which amounts to 4,925 test instances. For each of these test molecules, we take the provided formula and query all seed structures from PubChem, excluding the true molecule. We run each search with a termination criterion of 7,500 calls to ICEBERG.

\subsection{ICEBERG training}
\subsubsection{NIST'20 benchmarking details}
Training of ICEBERG on the NIST'20 reference library followed the same protocol as described in ref. \cite{Wang2025-tc}, where instructions to reproduce the training can be found. We utilize the model trained with random seed=1 and the random split for all NIST'20 experiments in the paper. The NIST'20 reference library dataset we curated covers 530,640 spectra covering 25,541 molecules collected only from Orbitrap instruments.

\begin{table}[ht]
\label{tab:nist20_adduct_training_exs}
\centering
\begin{tabular}{c|c}
\toprule
Adduct type & \# Training spectra \\
\midrule
{[M+H]\textsuperscript{+}} & 214331 \\
{[M\textminus H]\textsuperscript{\textminus}} & 89875 \\
{[M\textminus H\textsubscript{2}O+H]\textsuperscript{+}} & 58426 \\
{[M+Na]\textsuperscript{+}}  & 25582 \\
{[M\textminus 2H\textsubscript{2}O+H]\textsuperscript{+}}  & 11396 \\
{[M\textminus H\textminus CO\textsubscript{2}]\textsuperscript{\textminus}} & 9504 \\
{[M\textminus H\textminus H\textsubscript{2}O]\textsuperscript{\textminus}} & 7720 \\
{[M+NH\textsubscript{3} +H]\textsuperscript{+}} & 6263 \\
{[M+Cl]\textsuperscript{\textminus}} & 6032 \\
{[M+K]\textsuperscript{+}} & 622 \\
\bottomrule
\end{tabular}
\caption{Training spectra counts by adduct type (random split, NIST'20 dataset)}
\end{table}

\subsubsection{MassSpecGym benchmarking details}
\label{sec:si_msg_benchmark_details}
The MassSpecGym benchmark \cite{Bushuiev2024-jj} consists of 231,104 spectra. It is restricted to only cover [M+H]\textsuperscript{+} and [M+Na]\textsuperscript{+} adducts, structures of 1000 Da or less in molecular weight, but features spectra collected from QToF instruments in addition to spectra collected from Orbitrap instruments. Therefore, besides the additional instrument type, the coverage of spectra are more limited in scope compared to NIST'20. The spectra are sourced from several public repositories, including GNPS, MoNA, and MassBank, and though there is extensive quality preprocessing, their curation from heterogeneous sources makes their quality more variable compared to those in the NIST'20 reference library. 

For the MassSpecGym dataset, each spectrum collected at a separate collision energy must be considered independently, amounting to 17,082 testing instances. As consistent with DiffMS, we excluded 65 instances that include elements that are not supported by DiffMS, prohibiting sample generation for experiments. 

\textbf{Training ICEBERG}: As there are both Orbitrap and QToF spectra present in the dataset, we first modify ICEBERG to support an instrument parameter in addition to adduct and collision energy. 
Approximately half (47\%) of this dataset does not include annotated collision energy, which could suggest they may be a combination of spectra run at several collision energies. Nonetheless, to overcome this limitation for compatibility with ICEBERG, which is collision-energy aware, we use the ICEBERG model trained on the NIST'20 Orbitrap data to annotate collision energy for the unannotated spectra. We do so by performing a retrieval sweep over a range of 0-150 eV collision energy and selecting the energy with the lowest entropy distance to the experimental spectra. Inference for this task comprised 2.5 million fragmentations, which took approximately two weeks to run on 3 NVIDIA A5000 GPUs. This step only provides pseudo-annotations of collision energies and does not introduce other knowledge learned from the NIST dataset; the concern of data leakage is minimal. We then use these imputed collision energies to complete the training set, and train a new ICEBERG-Generate model and a ICEBERG-Score model. We train the ICEBERG-Score model with a spectral cosine loss instead of a spectral entropy loss as used by the NIST'20 ICEBERG setting. The weights for this model are available here: \url{doi:10.5281/zenodo.18502041}.

\textbf{DiffMS as a source of seed structures}: To collect DiffMS samples, 100 samples were collected for each of the spectra in the MassSpecGym dataset and deduplicated. Code and weights are used as released here: \href{https://github.com/coleygroup/DiffMS}{https://github.com/coleygroup/DiffMS}. Any fragmented structures are removed from the samples. Since DiffMS is heavy-atom constrained, samples may have an inconsistent number of hydrogens; only structures that match in the full formula are kept. If the true structure is present in these samples, it is not removed. Collecting all structure samples took approximately one week on 6 NVIDIA H100 GPUs. 

\subsection{Runtime}
For the NIST'20 experiments, 3 generations (approximately equivalent to 800 calls) takes 4 minutes per test instance; the full search (7,500 calls) takes approximately 20 minutes on average to complete. The spectrum simulation component of the method is GPU-compatible. To accelerate both generation and spectrum simulation, experiments were performed on 72-core CPU nodes that dispatched spectrum simulation of ICEBERG to another, GPU-rich node (NVIDIA H200, H100, or 2080Ti GPUs). We adopt a similar strategy for the MassSpecGym experiments, whose full searches (1,500 calls) take on average 3 minutes to run. As a reference point, on an NVIDIA H100 GPU, ICEBERG can score 500 molecules with coverage of 3 collision energies in 30 seconds.

\subsection{Parameter choice}
\subsubsection{NIST evaluation parameters}
\begin{tabular}{| m{10em} | m{17em} |c |}
    \hline
    Parameter & Parameter description & Parameter value \\ 
    \hline
    Oracle call cap & Number of allowed calls to ICEBERG before FOAM terminates & 7,500 \\  
    \hline
    Offspring size & Number of maximal new candidates generated at each iteration & 600 \\
    \hline 
    Population size & Size of candidate pool following non-dominated selection & 200 \\
    \hline 
    Maximum crossover attempts & Number of times crossover is attempted before failing & 10 \\
    \hline 
    Mutation rate & Rate of mutation to offspring made following crossover & 0.20 \\
    \hline 
    Truncation upper minima threshold & Similarity threshold above which if a sufficient number of molecules are present in the population, truncation will occur & 0.4 \\
    \hline
    Truncation percent required & Percentage of molecules that must be above the upper minima threshold for truncation to occur & 0.20 \\
    \hline
    Truncation cutoff threshold & Similarity below which any molecules are removed from the population if truncation occurs & 0.35 \\
    \hline 
    ICEBERG batch size & Number of inputs (each SMILES, duplicated by collision energy) passed to ICEBERG & 32 \\
    \hline 
    Number of ICEBERG workers & Copies of ICEBERG loaded onto a single 32GB GPU for add'l parallelism & 8 \\
 \hline 
\end{tabular}
\subsection{Parameter choice}
\subsubsection{MassSpecGym evaluation parameters}
\begin{tabular}{| m{10em} | m{17em} |c |}
    \toprule
    Parameter & Parameter description & Parameter value \\ 
    \midrule
    Oracle call cap & Number of allowed calls to ICEBERG before FOAM terminates & 1,500 \\  
    \hline
    Offspring size & Number of maximal new candidates generated at each iteration & 600 \\
    \hline 
    Population size & Size of candidate pool following non-dominated selection & 200 \\
    \hline 
    Maximum crossover attempts & Number of times crossover is attempted before failing & 10 \\
    \hline 
    Mutation rate & Rate of mutation to offspring made following crossover & 0.20 \\
    \hline 
    Truncation upper minima threshold & Similarity threshold above which if a sufficient number of molecules are present in the population, truncation will occur & 0.4 \\
    \hline
    Truncation percent required & Percentage of molecules that must be above the upper minima threshold for truncation to occur & 0.20 \\
    \hline
    Truncation cutoff threshold & Similarity below which any molecules are removed from the population if truncation occurs & 0.35 \\
    \hline 
    ICEBERG batch size & Number of inputs (each SMILES, duplicated by collision energy) passed to ICEBERG & 32 \\
    \hline 
    Number of ICEBERG workers & Copies of ICEBERG loaded onto a single 32GB GPU for add'l parallelism & 8 \\
    \hline
    Number of DiffMS samples collected & Number of initial seed structure proposals collected from DiffMS (spectrum-conditioned generation) & 100 \\
 \bottomrule
\end{tabular}

\subsection{Additional evaluations}
\begin{figure}[h!]
  \centering
  
  \includegraphics[width=\linewidth]{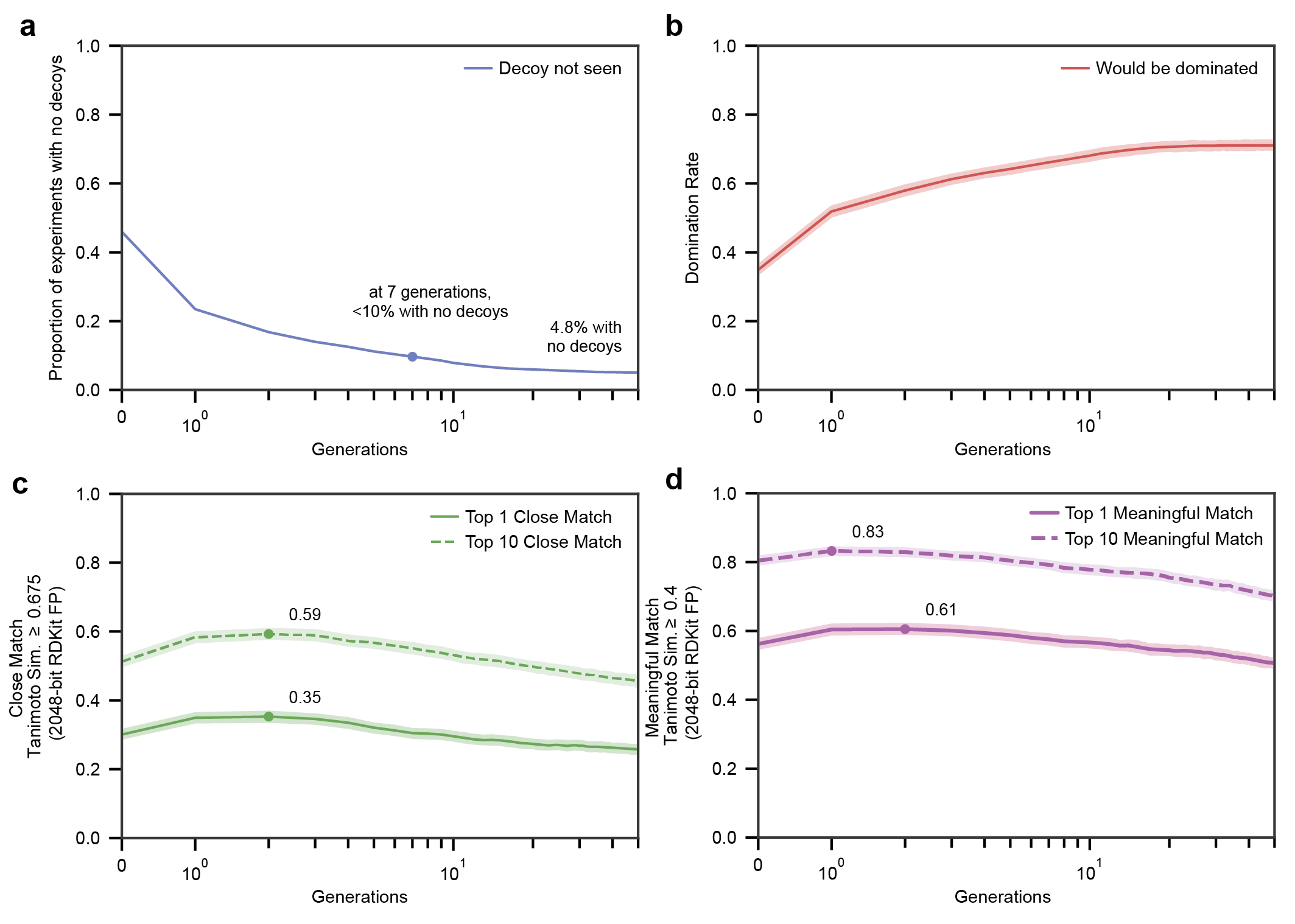}
  
  \caption{Additional evaluation of FOAM performance over time on NIST'20. (a) Proportion of experiments with no decoys observed, over course of trajectory. (b) Whether true structure would be dominated by population. (c) Following analytical chemist defined estimations of structural utility~\cite{Butler2023-pt}, Close match proportions in the top 1 and top 10 candidate sets, as well as (d) Meaningful match proportions in the top 1 and top 10 candidate sets. }
  \label{fig:nist20_decoys_ms2match}
  
\end{figure}

Beyond the evaluations provided, we additionally track the generative dynamics of FOAM's searches on the NIST'20 test set. In \ref{fig:nist20_decoys_ms2match}a, we first track what percentage of examples encounter a molecule with equal or higher predicted spectral similarity to the experimental spectrum than that of the true molecule. This plot highlights that by the end of the searches, only 4.8\% of searches have not yet encountered a decoy structure with equivalent or higher structural similarity. In  \ref{fig:nist20_decoys_ms2match}b, we show where the true molecule, whether present or not, would be dominated by the existing population (i.e, if FOAM were to produce the true molecule, the true molecule would have a Pareto rank greater than 1). This curve plateaus at around 70\%, suggesting that about 30\% of the instances have produced final populations where the true molecule would still be considered Pareto optimal.  In  \ref{fig:nist20_decoys_ms2match}c and d, we utilize the Close Match and Meaningful Match definitions from \cite{Butler2023-pt} to review how relevant our structures might be valuable to domain experts. Using RDKit topological fingerprints, suggestions are considered close matches if they have a Tanimoto similarity of 0.675 or higher, and meaningful matches if they have a Tanimoto similarity of 0.4 or higher. 

\begin{figure}[h!]
  \centering
  \includegraphics[width=0.7\linewidth]{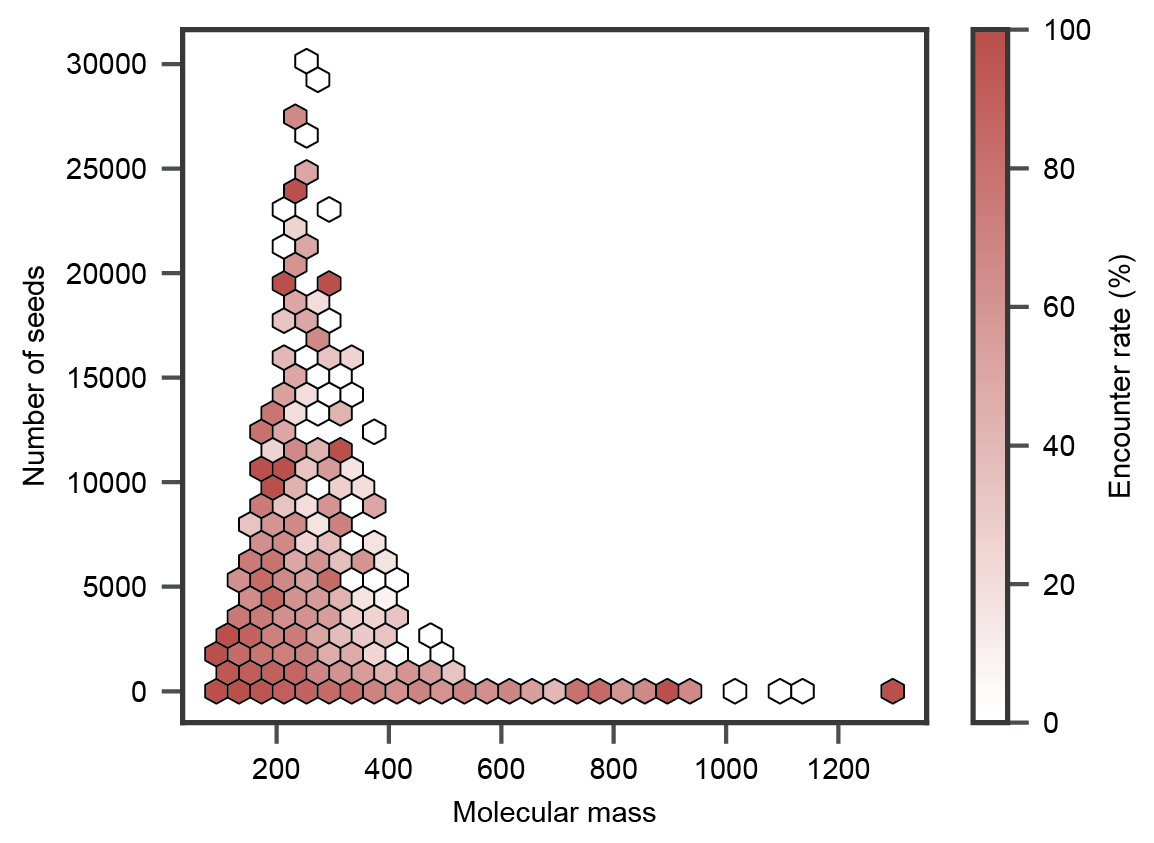}
  \caption{Distribution of count of seed structures from PubChem and mass versus average encounter rate observed for test spectra in the NIST'20 dataset.}
  \label{fig:SI_seeds_v_mass_spread}
\end{figure}
\begin{figure}[h!]
  \centering
  \includegraphics[width=0.8\linewidth]{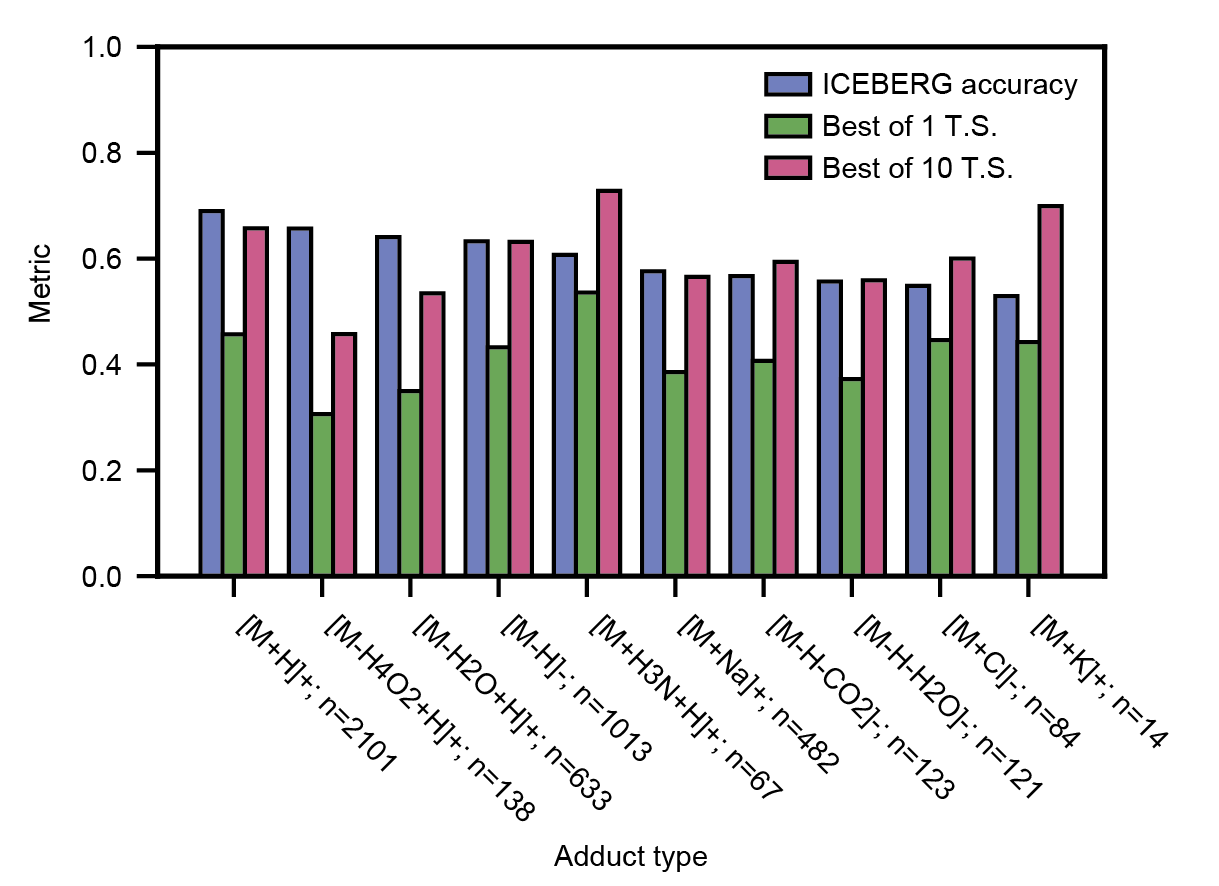}
  \caption{ICEBERG accuracy and structural similarity (T.S.: Tanimoto similarity) of the Best of 1 and Best of 10 structures, grouped by adduct. Test spectra counts for each adduct are denoted with each x-axis tick label.}
  \label{fig:si_nist20_by_adducts}
\end{figure}
\FloatBarrier
\clearpage
\subsubsection{Review of candidate sets after longer generation count (NIST'20)}
\begin{figure}[h!]
  \centering
  \includegraphics[width=0.44\linewidth]{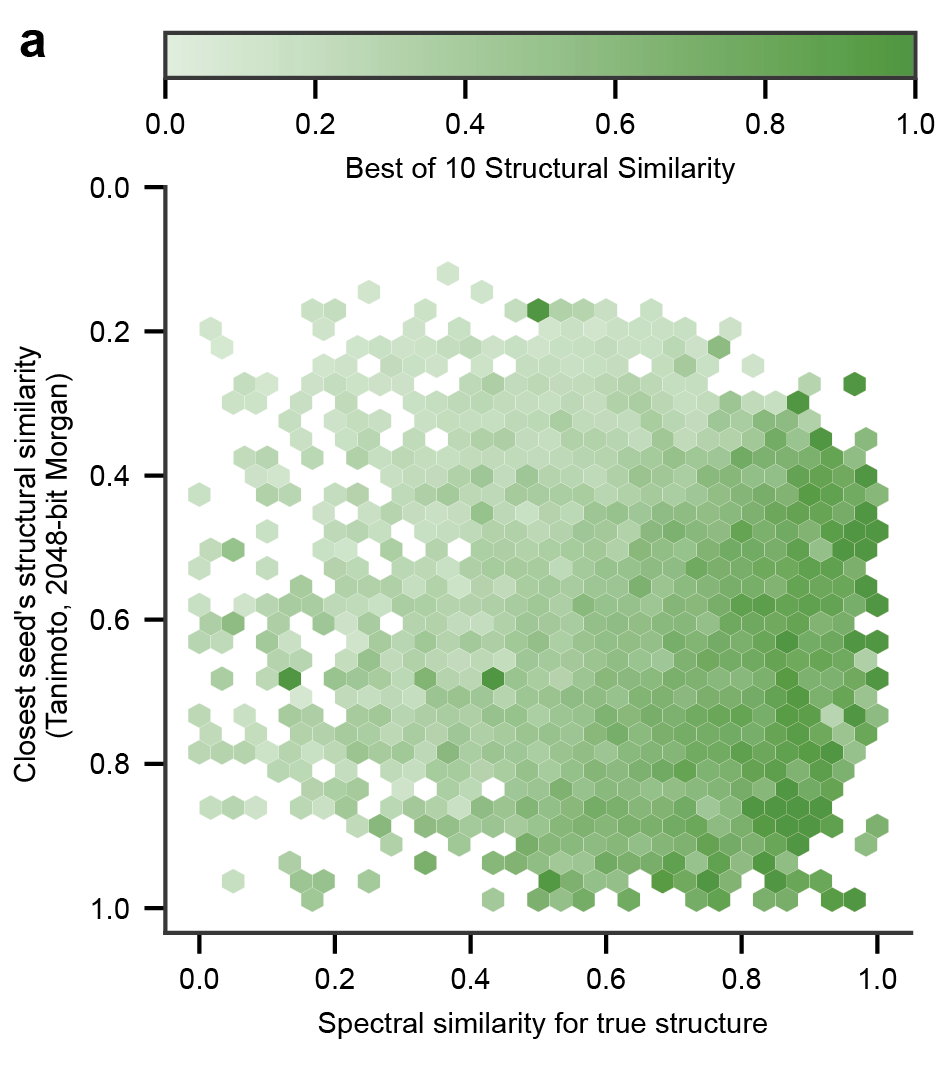}
  \caption{(a) Distribution of structural similarities of best candidates in the top 10 versus predicted spectral similarity to the experimental spectrum for the true structure (x-axis) and closest seed's structural similarity (y-axis), but after FOAM has run for 25 generations instead of 3 as in Figure \ref{fig:limitations}a. As such, this heatmap displays markedly right-shifted saturation, reflecting the ultimate importance of having an accurate objective measure. }
  \label{fig:25gen_heatmap}
\end{figure}
\FloatBarrier

\subsection{MassSpecGym evaluation results}
One clear differentiating factor within the MassSpecGym dataset is that almost half of the spectra are missing a collision energy annotation, indicating possibly merged spectra or some other modification to the data. On these affected spectra, the predicted spectral similarity of the true molecules associated with these spectra are on average considerably lower (0.193) than that of the spectra with known collision energies, either in the MassSpecGym (0.427) or NIST'20 datasets (0.648) (Extended Data Figure~\ref{fig:msg_evals}d). This translates to a poorer ability to rank the true molecule in the top 10, with FOAM only being able to find a match in the top 10 4.7\% of the time, compared to 14.0\% on the unaffected MassSpecGym spectra (Extended Data Figure~\ref{fig:msg_evals}d). Additionally, with successive iterations, this less accurate simulator ranks progressively more decoy structures, which have higher simulated spectral similarity but seem to lose structural similarity to the true structure. 

\end{document}
\typeout{get arXiv to do 4 passes: Label(s) may have changed. Rerun}